\documentclass{bmcart}

\usepackage{amsthm,amsmath}
\RequirePackage[authoryear]{natbib}
\RequirePackage{hyperref}
\usepackage[utf8]{inputenc} 

\usepackage{graphicx} 

\startlocaldefs

\usepackage{multirow}
\usepackage{array}
\usepackage{comment}
\usepackage{mathrsfs}

\newcolumntype{P}[1]{>{\centering\arraybackslash}p{#1}}
\newcolumntype{L}[1]{>{\arraybackslash}p{#1}}

\newcommand{\ns}{\mkern-0.5mu}
\newcommand{\nss}{\mkern-1.5mu}

\newcommand{\circlenumber}[1]{\raisebox{.5pt}{\textcircled{\raisebox{-.9pt} {#1}}}}

\endlocaldefs

\begin{document}

\begin{frontmatter}

\begin{fmbox}
\dochead{Research}


\title{Teachers' perception of Jupyter and R Shiny \\as digital tools for open education and science}


\author[
   addressref={aff1},                   
   corref={aff1},                       
   noteref={},                        
   email={jozef.hanc@upjs.sk}   
]{\inits{JH}\fnm{Jozef} \snm{Han\v{c}}}
\author[
   addressref={aff1},
   noteref={},
   email={peter.strauch@student.upjs.sk}
]{\inits{PS}\fnm{Peter} \snm{\v{S}trauch}}
\author[
addressref={aff1},
noteref={},
email={eva.pankova@student.upjs.sk}
]{\inits{PS}\fnm{Eva} \snm{Pa\v{n}kov\'a}}
\author[
addressref={aff2},
 noteref={},                        
email={martina.hancova@upjs.sk}
]{\inits{MH}\fnm{Martina} \snm{Han\v{c}ov\'a}}


\address[id=aff1]{
  \orgname{Institute of Physics, Faculty of Science, Pavol Jozef \v{S}af\'{a}rik University}, 
  \city{Ko\v{s}ice},                              
  \cny{Slovakia}                                    
}
\address[id=aff2]{%
  \orgname{Institute of Mathematics, Faculty of Science, Pavol Jozef \v{S}af\'{a}rik University}, 
\city{Ko\v{s}ice},                              
\cny{Slovakia}                                    
}


\begin{artnotes}
\note[id=n1]{Equal contributor}  
\end{artnotes}

\end{fmbox}


\begin{abstractbox}
	
	\begin{abstract} 
		During the last ten years advances in open-source digital technology, used especially by data 
		science, led to very accessible ways how to obtain, store, process, analyze or share data in almost 
		every human activity. Data science tools bring not only transparency, accessibility, and reproducibility in open science, but also give benefits in open education as learning tools for improving effectiveness of instruction.
		
		Together with our pedagogical introduction and review of Jupyter as an interactive multimedia learning tool we present our three-years long research in the framework of a complex mixed-methods approach which examines physics teachers' perception of Jupyter technology in three groups: Ph.D. candidates in physics education research (PER) ($N = 9$), pre-service physics teachers ($N = 33$) and in-service physics teachers ($N = 40$).
		
		Despite the fact that open-source Jupyter notebooks are natural and easy as email or web, the results suggest that in-service teachers are not prepared for Jupyter technology and open analysis, but positively accept open education data presented via another open-source data science tool, R Shiny interactive web application, as an important form of immediate feedback and learning
		about the quality of their instruction. 
		
		Simultaneously our instruction results in the frame of the Flipped Learning also indicate that young beginning PER researchers and pre-service physics teachers can master key digital skills to work with 
		Jupyter technology appreciating its big impact on their learning, data and statistical 
		literacy or professional development.
		
		All results support the ongoing worldwide effort to implement Jupyter in traditional education as a promising free open-source interactive learning tool to foster learning process, especially for the upcoming young generation.
		
		%
	\end{abstract}
	
	
	
	
\end{abstractbox}
%

\end{frontmatter}




\section*{Introduction}

During the last decade, new digital technologies such as mobile devices, cloud infrastructure, open data, artificial intelligence, decentralized and social networks \citep{downes_look_2019} caused that we live in the digital and data-intensive age. Ninety percent of all data has been created in the last two years \citep{boaler_data_2020, marr_how_2018}. Data touches all aspects of our lives. The world economy, our jobs, our health, our environment and our roles as citizens increasingly depend on the knowledge, skills and technology required to work with, understand, and effectively use data \citep{boisvert_building_2016}.

In April 2020 GitHub \citep{github_inc_github_2020, warner_thank_2018}, the world largest web cloud platform for storing, social coding and collaborating on any open code or digital content, reached a further milestone, 50 million developers working on over 100 million repositories --- 3,000 times more than after its starting year 2008. 

Simultaneously, during last years open-source software and data science tools,  namely programming languages Python and R together with environments for working with them --- especially Jupyter and RStudio, have conquered the data science world \citep{kaggle_kaggles_2020} providing everybody free, open, revolutionary and very accessible ways how to gather, store, process, analyze, present or share data in almost every human activity. 

Speaking about science, in the near future, the European Open Science Cloud (EOSC, \href{https://eosc-portal.eu/}{eosc-portal.eu}), officially launched at the end of 2018, will offer to millions of European researchers, professionals, but also university students, in science, technology, the humanities and social sciences a virtual environment with open and seamless services for storage, management, analysis and re-use of open research data, interoperable among all scientific domains or EU member states. By federating, the EOSC will connect existing and being built scientific data infrastructures, currently dispersed and isolated across disciplines and borders of EU states. One of the key and easily accessible open digital research environments in EOSC (see e.g. the infrastructure EOSC project OpenDreamKit, 2015-2019, \href{https://opendreamkit.org/}{opendreamkit.org}) should become the mentioned technology of Jupyter Notebooks. The highest ambition of the EOSC is to change the way we do science, to open it.

As for education, in 2015 from US a worldwide \textit {Call for Action to promote data literacy} swept the world \citep{edc_oceans_of_data_institute_building_2015,boisvert_building_2016}. Its signatories call for ``a revolution in education, placing data literacy at its core, integrated throughout K-16 education nationwide and around the world. By enabling learners to use data more effectively, we prepare them to make better decisions and to lead more secure, better-informed and productive lives.''

On February 2020, Jo Boaler, a professor of mathematics education at Stanford, invited a group of fifty mathematicians, data scientists, teachers and education policy leaders to start a movement, a YouCubed initiative, which will modernize the K-12 math curriculum and will prepare high school students to the data age \citep{spector_bringing_2020, boaler_data_2020}.

\subsection*{Jupyter in science and education}

One of the current main data science tools for open science is Jupyter technology of interactive multimedia notebooks --- shortly and officially called \textit{Jupyter Notebooks}. From the viewpoint of users, the technology is natural and easy as email or web. It was released in 2014 in the frame of \textit{Project Jupyter} \citep[\href{https://jupyter.org/}{jupyter.org},][]{kluyver_jupyter_2016,projectjupyter_binder_2018} as a free, open alternative to the well-known Mathematica Notebooks from commercial software Wolfram Mathematica (\href{https://www.wolfram.com}{wolfram.com}).  
As for the technology platform, Project Jupyter evolved from IPython, a personal side project of a Jupyter-project co-founder and former nuclear physicist, Fernando Peréz. 

Jupyter is an open source, free interactive web computing environment, accessible through any modern web browser, that enables users to use, modify or create interactive multimedia web documents which mix live code (over 100 computer languages  with a focus on Python, called kernels), interactive computations, equations, simulations, plots, images, narrative texts, annotations, audios or videos. 

\begin{figure}[h!]
\includegraphics[width=0.9\linewidth]{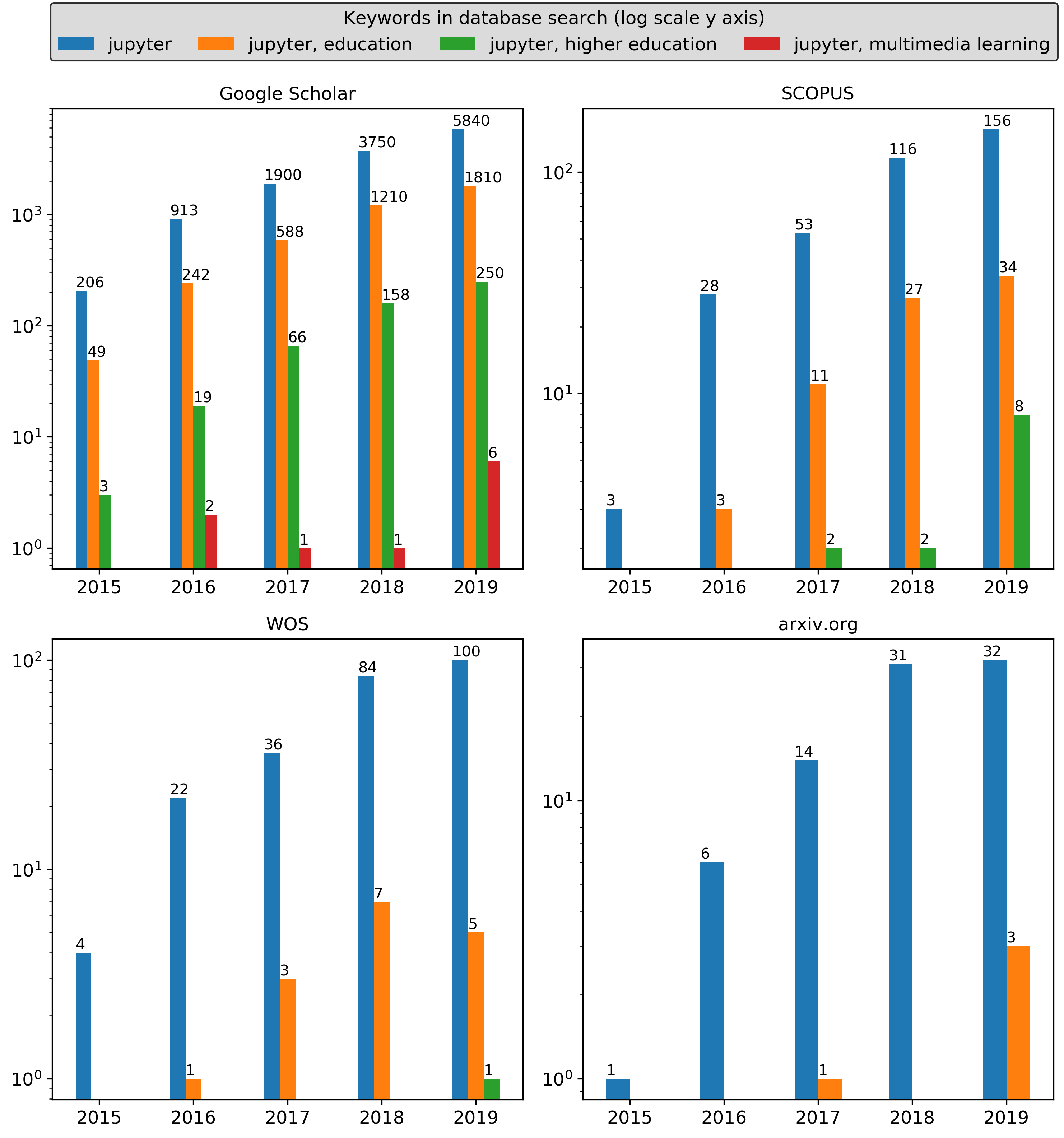}
\caption{Annual numbers of scientific publications dealing with Jupyter in four scholar databeses during 2015-2019. Our review as bar plots was effectively created using \textit{Scientific Python} in a Jupyter notebook (data analysis library \textit{pandas}).}
\label{fig:JupyterLiterature}
\end{figure}

Over the last 5 years, Jupyter Notebooks have become the most widely used as an environment for performing scientific calculations, open analysis, data processing and scientific reporting \citep{frederickson_ranking_2019}. Data and analysis of the recent Nobel prize-winning detections of gravitational waves (2015), predicted by Einstein in 1915, were released as a Jupyter Notebook. At the same time we are witnesses of an explosion of scientific and technical articles dealing with Jupyter's application in science, technology and education (Fig.~\ref{fig:JupyterLiterature}).

As we can see in hundreds of scholarly papers connected to education (Fig.~\ref{fig:JupyterLiterature}), e.g. \cite{weiss_scientific_2017} -- chemistry, \cite{koehler_interactive_2018} -- mathematics, \cite{odden_physics_2019} -- physics, \cite{wright_why_2020} -- biology, \cite{cardoso_using_2019} -- engineering, Jupyter is widely applied in open STEM and STEAM education \citep{khine_steam_2019}, supporting data literacy and including also programming, statistics, data science, cognitive science, computer science, machine learning, digital humanities, scientific computation or robotics \citep{barba_teaching_2019}.

Today, Jupyter notebooks, as living interactive ``storytelling'' documents, is the technology behind many innovative educational programs and also became platform-of-choice for tutorials, workshops, online lessons, and even books. One of the best elaborated book examples is an open GitHub handbook \textit{Teaching and Learning with Jupyter} \citep{barba_teaching_2019} covering topics like why and how to use Jupyter in education, pedagogical instruction designs and case studies elicited from authors' real experience or technical details of Jupyter implementation in practice. Another example is Coursera (\href{https://www.coursera.org/}{coursera.org}), a world-wide MOOC platform, offering many free courses and guided projects using Jupyter as a key learning tool.

The revolutionary significance of Jupyter for open science and education also became the subject of philosophical reflections, e.g. from a writer and programmer J. Sommers \citep{somers_scientific_2018} or a Nobel prize laureate P. Romer \citep{romer_jupyter_2018} or a Canadian philosopher and open education visionary S. Downes \citep{downes_look_2019}.

\subsection*{Jupyter as a tool of interactive teaching methods}

Finally in this introductory section, we would like to point out three important, but less known, Jupyter features in connection with three modern interactive instruction approaches which are ``designed to promote conceptual understanding through interactive engagement of students in heads-on (always) and hands-on (usually) activities which yield immediate feedback through discussion with peers and/or instructors'' \citep{hake_interactive-engagement_1998,redish_oersted_2014, pankova_practical_2016,fraser_teaching_2014}.

\paragraph{Inquiry based science education.} Jupyter allows educators to narrate a ``conversation'' between the student, concepts and data with goals such as building a model, carrying out a virtual experiment (simulation) or visualizing any data or process, all with or without programming. Pedagogically, such activities can be designed in the sense of IBSE \citep{heering_innovative_2012, constantinou_what_2018}.

\begin{figure}[!htpb]
\includegraphics[width=0.975\linewidth]{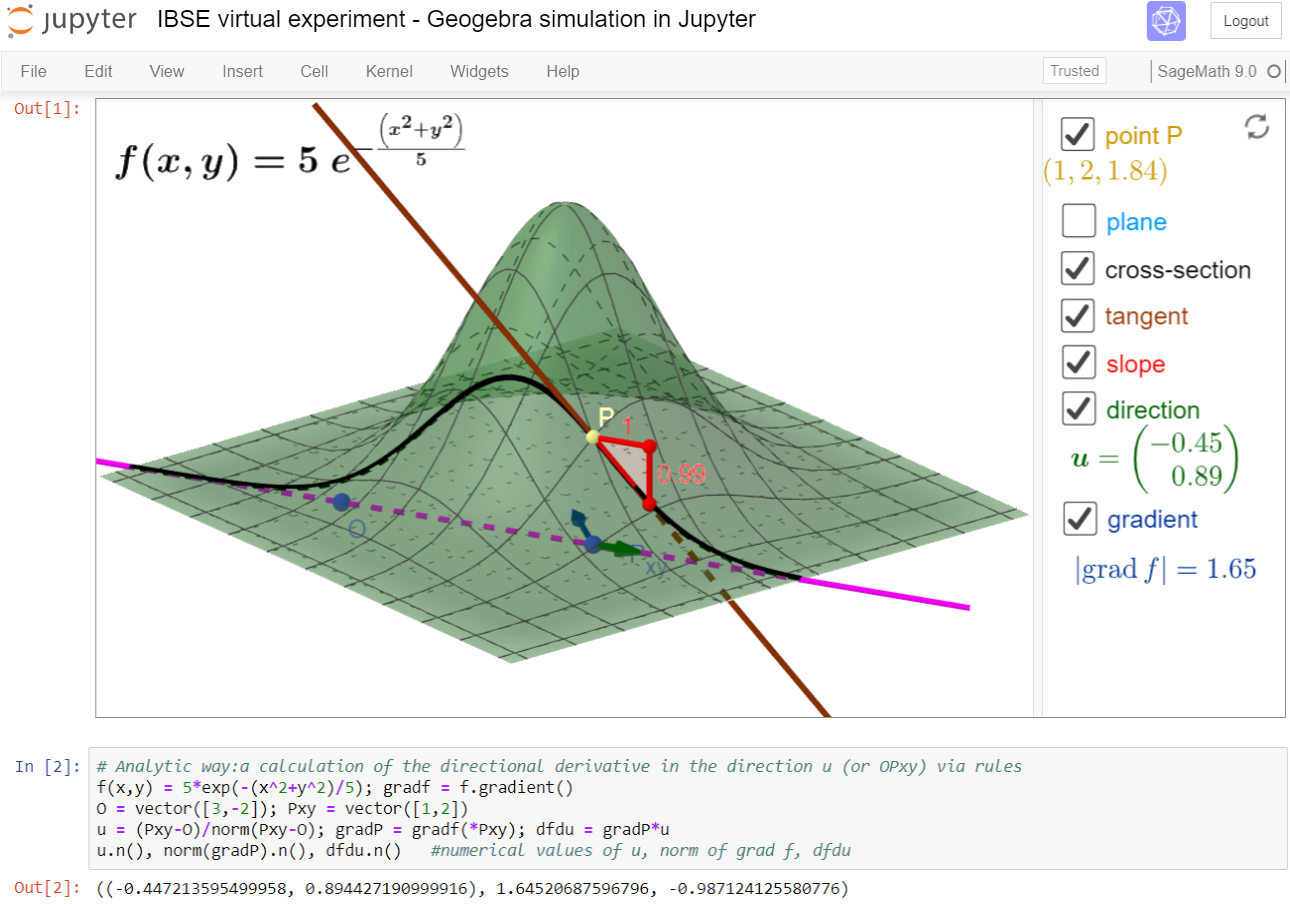}
\caption{A Jupyter notebook example with an embedded interactive 3D Geogebra simulation and a SageMath code which can be applied as part of virtual experimentation in the frame of IBSE.}
\label{fig:JupyterGeogebra}
\end{figure}

To be more specific, Jupyter has the so-called \textit{widget functionality} which provides the notebook user access to slide bars, toggle buttons or text-boxes. Such elements can hide the code and allow us to create a notebook app or simulation with a primary goal to explore or visualize model, computation or data. Another very useful feature of Jupyter Notebook as a multimedia web document, is \textit{embedding} functionality allowing to embed any available digital content from Web (via IPython command \texttt{IFrame} or \texttt{HTML}). Regarding IBSE virtual experimentation and modeling, authors of the paper embed own or available interactive cloud Geogebra simulations (see our example in Fig.~\ref{fig:JupyterGeogebra} )\footnote{\footnotesize GeoGebra \citep[\href{https://www.geogebra.org/}{geogebra.org},][]{hohenwarter_geogebra_2018} is the leading dynamic mathematics software for STEM education allowing to create own interactive simulations without programming knowledge. Now it is offering over 1 million free activities, simulations, exercises, lessons and games for supporting STEM education and innovations worldwide.} and mix with Jupyter calculations in their university math or physics subjects \citep{bu_model-centered_2011,hall_mathematical_2016, hanc_geogebra_2011}.

\paragraph{Peer instruction and Question driven instruction.} In physics education, Peer instruction \citep{simkins_just--time_2009, crouch_peer_2001,fraser_teaching_2014} and Question driven instruction 
\citep{banks_question_2006,beatty_technology-enhanced_2009} are two well-known and very similar interactive methods\footnote{\footnotesize Both methods were established at the beginning of 1990s during the testing clickers in real school conditions at US universities and they are based on repeated peer-instruction or question cycles: (1) posing a question (problem) by the instructor; (2) small-group work of students on solutions – peer instruction; (3) collecting answers of students by e-voting; (4) displaying the answers without revealing the correct answer; (5) class-wide discussion; (6) closure (e.g. summarizing the key points or giving an explanation).} which promote student active learning based on constructivism ideas, formative assessment and cooperative learning. The key technological element of PI or QDI is e-voting which can be realized effectively by virtual clickers  (e.g. via cloud service \href{https://www.polleverywhere.com/}{polleverywhere.com}) and embedded in a Jupyter notebook or it can be also done very easily in Jupyter notebooks using \textit{Activity} extension \citep{barba_teaching_2019, blank_calystometakernel_2020}.

\paragraph{Flipped learning.} Interactive methods like IBSE, PI and QDI can be integral parts of group space activities\footnote{\footnotesize All these methods are regularly used by the authors during group space activities in flipped math and physics at secondary school and college levels \citep{hanc_application_2013,pankova_practical_2016,pankova_teaching_2019}.} in the flipped learning \citep{talbert_flipped_2017, nouri_flipped_2016, bergmann_flip_2012}, which is according to Talbert  ``a pedagogical framework in which the first contact with new concepts moves from group learning to individual learning space in the form of structured activity, and the
resulting group space is transformed into a dynamic, interactive learning environment where the educator guides students as they apply concepts and engage creatively in subject matter''.
As for a screencast technology for creating teaching materials used in the first-contact, pre-class work of students (their home preparation), Jupyter offers \textit{Graffiti} extension \citep{downes_look_2019, kessler_jupytergraffiti_2020}. This Jupyter extension allows to create interactive screencasts or ``live videos'' inside Jupyter Notebooks that student can watch and pause any time. During any pause, student can interactively ``play'' with the recorded instructor's work and combine it with own work and ideas. 

\newpage
\section*{Research purpose and design}

Typically, many of the found scholar papers (Fig.~1) are informal case studies, which are generally based on best teaching practices and wisdom of authors. According to a distinguished cognitive psychologist Richard Mayer \citep{mayer_applying_2008} such applied approach in instruction usually leads to a set of empirical-based design principles but with limited applicability since they are not directly or weakly connected to the cognitive or psychological theory of learning which would provide solid grounds how or why they work.

From the perspective of cognitive learning theory \citep{mayer_cambridge_2014}, Jupyter notebook as a learning tool belongs to multimedia (online) which combines words and pictures in static and dynamic form \cite{mayer_thirty_2019}. Simultaneously Jupyter notebooks are key elements of computer-based learning or online learning \citep{mayer_thirty_2019} when instruction is delivered on a digital device with intention to support learning. 

It means that any successful application of Jupyter with general applicability must respect the cognitive theory of multimedia and online learning whose principles and processes were discovered during last thirty years and can be found in \cite{mayer_cognitive_2014, mayer_thirty_2019}\footnote{\footnotesize It seems that The Cambridge Handbook of Multimedia Learning \citep{mayer_cambridge_2014} is the only comprehensive research-based reference monograph on multimedia and online learning.}. Our basic review of research literature in Fig.~\ref{fig:JupyterLiterature} demonstrates that up-to-date only a few scholar papers refer or try to apply or connect their educational research in Jupyter application with the mentioned cognitive principles. Other caveats of the considered empirical research are connected to the fact that the current cognitive theories of multimedia and online learning themselves would benefit from stronger incorporation of affect, motivation and metacognition where research is in its infancy \citep{mayer_thirty_2019}.

\paragraph*{Research purpose.}Therefore the purpose of our research was to explore the teacher's perception of Jupyter as a digital technology during and after their higher education with the focus on affective aspects of learning. We examined teacher's perceptions from two viewpoints. The first viewpoint concentrated on the teacher's personal experience and reflections on how the use of Jupyter contributed to the understanding of learning content, what self-progress and how transferable were own application of technology in other context. The second viewpoint focused on the teacher's overall, comprehensive impression and satisfaction from user, ``customer'' experience representing affective aspects --- own emotions and attitudes when experiencing Jupyter. Particularly, we were interested in feelings of Jupyter \textit{attractiveness}, \textit{difficulty} to get familiar with it, \textit{efficiency} during work with it, \textit{motivation} in using it and \textit{capturing} user's attention.

\paragraph*{Research design.} Our three-years long research was set in the framework of a complex mixed-methods approach with the convergent design where we combined results of three parallel research studies providing all our available complementary sources of quantitative and qualitative data, in order to best understand the research problem \citep{creswell_designing_2017,johnson_educational_2016}. We describe the studies conducted from June 2017 to June 2020 in the next three sections when we also address in detail for each study: background, context, particular design, participants, procedure, methods and results with the corresponding discussion. 

\paragraph{Data collection and analysis tools.}
Diagnostic, data collection and analysis methods will be described in the given studies. As for technology, all statistical analysis in all three studies with data visualization, manipulation, and processing were carried out in open-source data science tools --- Jupyter Notebooks with kernels (1) scientific Python \citep[][SciPy,]{jones_scipy_2001,oliphant_python_2007} and (2) R \citep{r_development_core_team_r:_2020} using free available Python libraries: \textit{numpy} \citep{walt_numpy_2011}, \textit{pandas} \citep{mckinney_data_2010}, \textit{matplotlib} \citep{hunter_matplotlib_2007} and R libraries: \textit{RcmdrMisc} \citep{fox_rcmdrmisc_2020}, \textit{sjstats} \citep{ludecke_sjstats_2020}, \textit{dplyr}, \textit{readxl}, \textit{scales}  \citep{wickham_dplyr_2020,wickham_readxl_2019,wickham_scales_2020}, \textit{cluster} \citep{maechler_cluster_2019}, \textit{factoextra} \citep{kassambara_factoextra_2020}, \textit{hmisc} \citep{harrelljr_hmisc_2020}, \textit{psychometric} \citep{fletcher_psychometric_2010}, \textit{repr} \citep{angerer_repr_2020} and \textit{shiny} \citep{chang_shiny:_2019}.

Our open data analysis in the form of Jupyter notebooks together with all used tool and data files are stored and freely available at one of our GitHub repository devoted to this paper \citep{hanc_jupyterper_2020-1} in the frame of our GitHub research project \textit{Jupyter in Physics Education and Research} \citep{hanc_jupyterper_2020}.

\section*{Study I: Ph.D. candidates in physics education research}

\vspace{6pt}

\subsection*{Background and context}

In 2017, during one of the last preparatory Ph.D. seminars before the thesis defense, we observed that  in their presentation only one of our six Ph.D. candidates in Physics Education Research (PER) has applied data science tool R in own data analysis. All of them were our first students who completed the course, taught by JH in the collaboration with MH from a statistical department at our university, where the main goal was the use of advanced statistical methods in the \textit{R Commander} \citep{fox_using_2016,fox_rcmdr_2020}, a simple point-and-click graphical interface for R. 

The finding was really disturbing since the course was pedagogically designed in the frame of flipped learning applying interactive and active-learning methods. It became a starting motive of our research to investigate what are the main reasons of the failure, why and how to change it. 

\subsection*{Participants}
The study involved PER graduate students ($N = 9$) with the master degree in teaching physics with other science subject, in that case math or biology. At our department of physics education the students were enrolled in the Ph.D. course \textit{Statistical  methods in Educational Research} where they experienced R with R Commander and Excel (up to 2017) or later R in Jupyter as a kernel. The left part of Table~\ref{tab:PhDstudents} in the results section shows sample characteristics.

\subsection*{Design and methods}

The qualitative case study had two sequential phases. In the first phase (May-June 2017), we retrospectively analyzed the use of statistical methods and digital technologies applied by 6 participants in their Ph.D. thesis or projects (Table~\ref{tab:PhDstudents}). Our research methods included the content analysis of Ph.D. theses (or Ph.D. project if the student did not yet complete the thesis) and a short qualitative interview containing two questions connected to 
key obstacles in using R with R commander and reasons for choosing an alternative (for the exact question wording see Appendix.) 

In the second phase (June 2017-June 2020), based on results of the first one, we decided for an intervention. In the same year (2017) we chose interactive Jupyter notebooks as the main tool for teaching and learning advanced statistical methods in PER. Moreover, three remaining participants, after completing the revised course, more regularly reported and discussed their progress in the application of chosen methods and tools. We interacted with them and observed them longitudinally during the whole period 2017-2020. 

\subsection*{Results and discussion}

The right part of Table~\ref{tab:PhDstudents} presents a summary of our content analysis focusing on applying advanced statistical methods, not only basic descriptive and inferential univariate statistics (column stats) and new digital technologies (cloud, special tools), not only Excel spreadsheets. Excel remained the main analysis tool and if necessary, students used special Excel add-ons, a trial version of XLSTAT \citep{addinsoft_xlstat_2020} and RSRPS \citep{zaiontz_real_2019}. Thanks to collaboration in her Ph.D. research, one student had access with professional statistical help to SPSS software \citep{ibmcorp_ibm_2015}, the very powerful commercial, but also expensive statistical software\footnote{\footnotesize In the first decade of this century, SPSS was one of the most popular in psychology, social sciences, market research, business and government \citep[e. SPSS]{salkind_encyclopedia_2010}. Today, it is still very widespread providing relatively easy access to modern and advanced statistical methods.}. One student also processed his pilot research data in online web service Data Explorer at PhysPort \cite[\href{https://www.physport.org/}{physport.org}]{mckagan_physport_2019}, a webpage of AAPT to empower physics faculty to use effective research-based physics teaching. 

\begin{table}[!h]
\renewcommand{\arraystretch}{1.3}
\caption{Using digital tools for data processing and analysis in PhD projects and thesis ($N=9$)} \label{tab:PhDstudents}
\begin{tabular}{clllr|llll}
	\hline
	& \multicolumn{4}{c|}{sample characteristics} & \multicolumn{4}{c}{content analysis results} \\
	phase &nick & g &      phd &  date &  stats & cloud? &         spreadsheets &          special sw\\
	\hline
	\multirow{6}{*}{I}&PhD1 &      F &   thesis &  2017 &  basic &  local &                Excel &                none \\
	&PhD2 &      F &   thesis &  2017 &  basic &  local &                Excel &                   R \\
	&PhD3 &      F &   thesis &  2017 &   adv. &  local &                Excel &                SPSS \\
	&PhD4 &      F &   thesis &  2017 &  basic &  local &                Excel &               RSRP \\
	&PhD8 &      M &  project &  2017 &  basic &  cloud &                Excel &            PhysPort \\
	&PhD5 &      F &   thesis &  2019 &   adv. &  local &                Excel &              XLSTAT \\
	\hline
	\multirow{4}{*}{II}&PhD9 &      M &  project &  2019 &  basic &  local &                Excel &                none \\
	&PhD6 &      F &   thesis &  2020 &   adv. &  cloud &  Excel,Google  &           R,Jupyter \\
	&\multirow{2}{*}{PhD7} &      \multirow{2}{*}{M} &   \multirow{2}{*}{thesis} &  \multirow{2}{*}{2020} & \multirow{2}{*}{adv.} &  \multirow{2}{*}{cloud} &  \multirow{2}{*}{Excel,Google } &  R,PhysPort, \\
	&     &        &          &       &        &        &                      &  Jupyter \\
	\hline
\end{tabular}

\end{table}

Regarding the qualitative interview here is a summary of main results representing participants responses with typical examples:

\vspace{6pt}
\begin{itemize}
\item \textit{onetime use}: ``Since I met R and R Commander only in this one-semester long course, and after one year of not using, I simply forgot it.'' 
\item \textit{installation problems}: ``I started to apply R but after reinstallation, I was not able to run it.''
\item \textit{difficult reproducibility}: ``During the course, I understood all the important things and everything looked so simple. But when it came time to analyze my data, I could not fit it to my data.''
\item \textit{steep learning curve}: ``I was trying really hard to use some R scripts from Web to apply more advanced methods, but it was difficult to understand.''
\item \textit{easier way}: ``Excel is for me still more simple and natural than R Commander.''; ``I found a great online tool [PhysPort] for my data and during real analysis I saw that I really did not need R for my Ph.D.'',
\item \textit{ignoring advanced methods}: ``My Ph.D. research did not need methods shown in the course.''
\end{itemize}
\vspace{6pt}

After our intervention situation was changed. Two of three Ph.D. candidates (the paper's coauthors -- PS, EP) fully relied on collecting, processing and analyzing data on Jupyter and R\footnote{\footnotesize The third one is planning to use data science tools in his research, but now from personal reasons he has a study break.}.  In thr previous Ph.D. works students used methods  and tools for analysis almost in a purely pragmatic way, only reproducing existing procedures. Now, thanks to better and ``more durable`` understanding which naturally resulted from working with Jupyter and its interactive and ``storytelling'' features, both students used advanced data visualization or analysis, even in new creative ways. 

\begin{figure}[!h]
\includegraphics[width=0.9\linewidth]{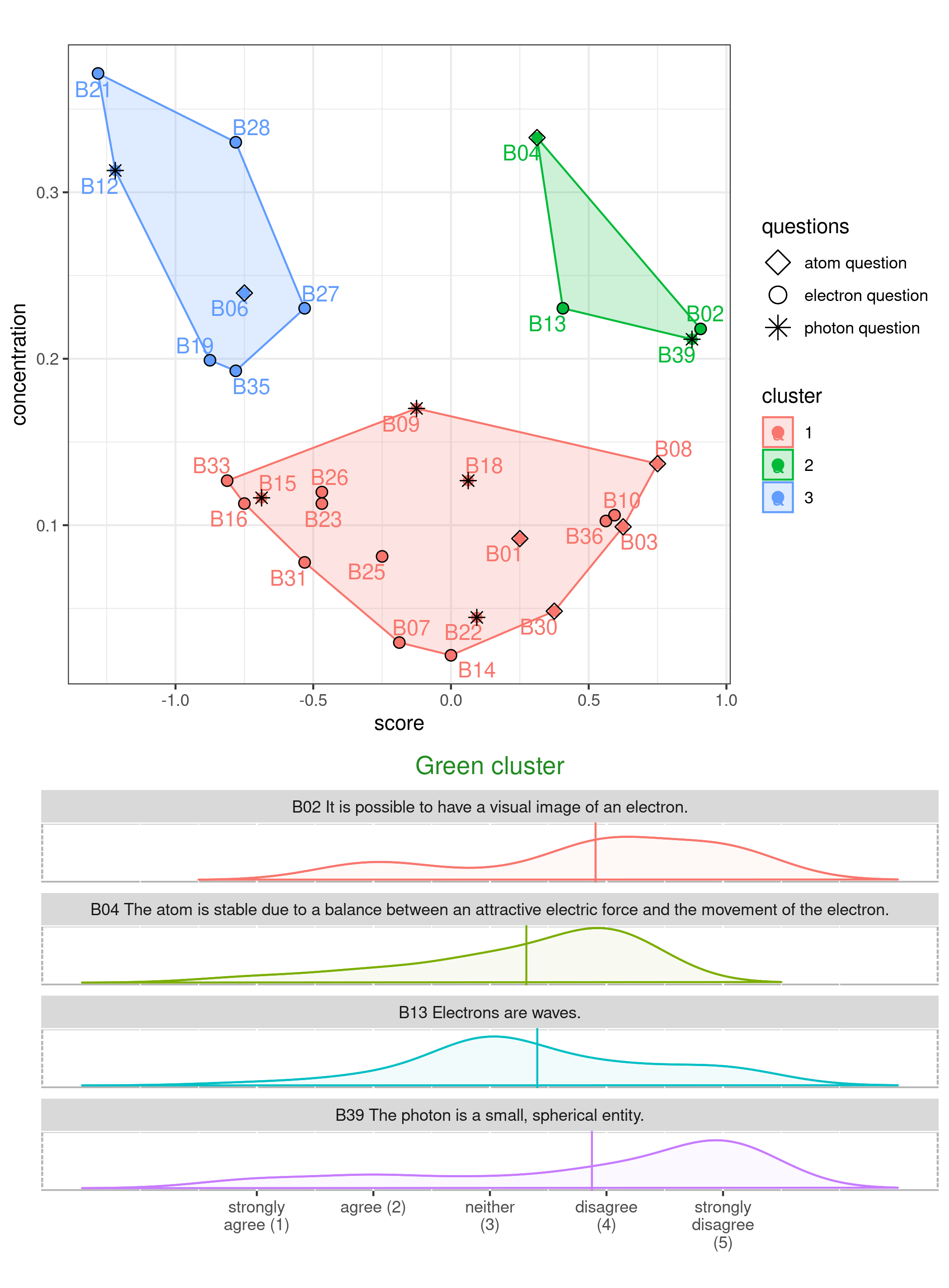}
\caption{An example of advanced multivariate analysis method using R in Jupyter: Hierarchical cluster analysis AGNES based on concentration factor \citep{bao_concentration_2001} and score computed and visualized using \textit{R software} (R packages \textit{cluster, factoextra}) supplemented by a continuous density estimation for the distribution of student's mental images behind quantum physics concepts (R package \textit{likert}) in the cluster.}
\label{fig:clusteplotscbig}
\end{figure}

One of such examples is illustrated in Fig.~\ref{fig:clusteplotscbig} from the Ph.D. thesis of EP \citep{pankova_flipped_2019}. We explain the corresponding details in the following discussion of the study results. The second example is presented in our parallel studies (Study III comes from the Ph.D. thesis of PS) dealing with a weighted benchmark analysis and plots of our data collected by UEQ (User Experience Questionnaire).


\begin{figure}[]
\includegraphics[width=0.9\linewidth]{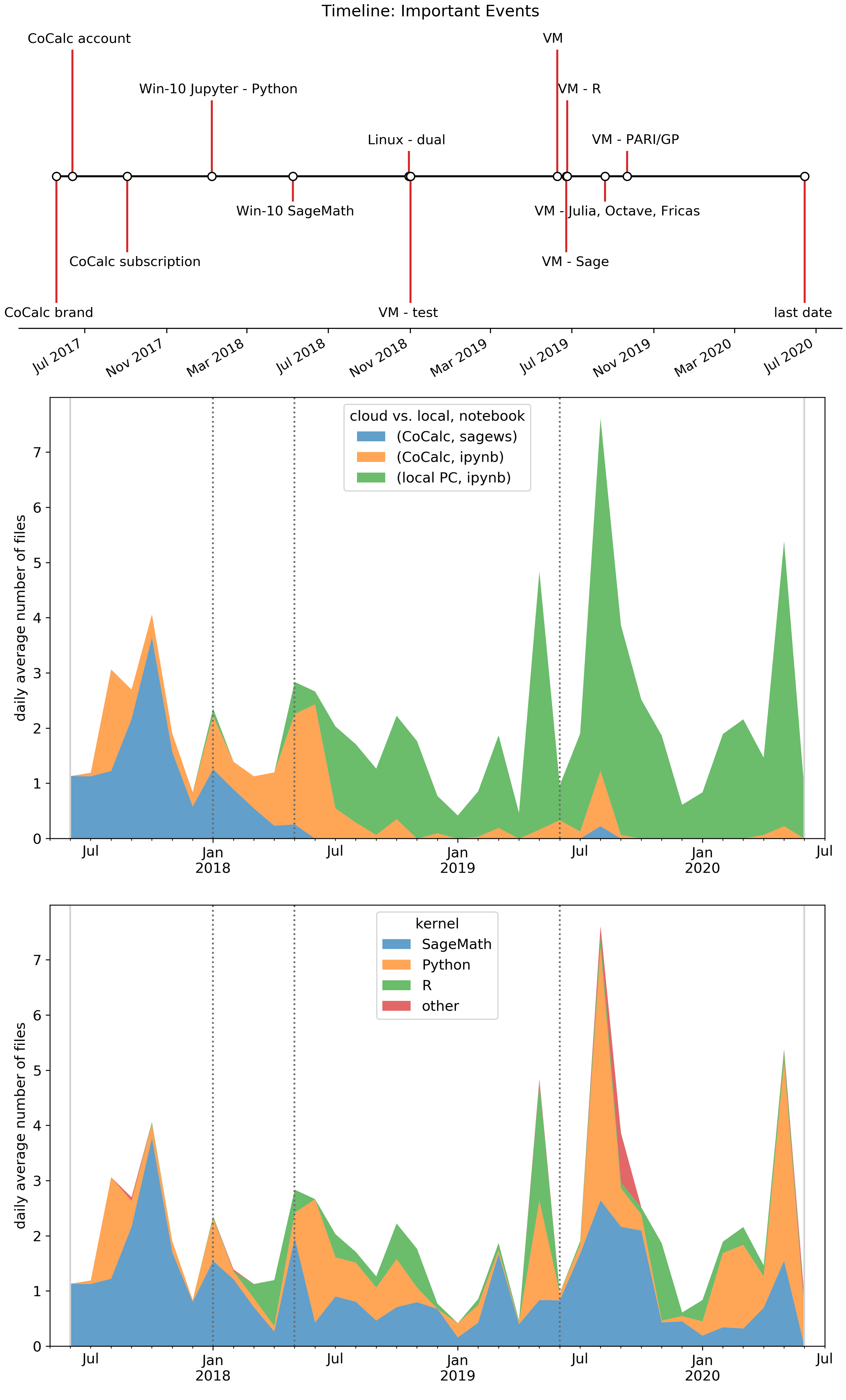}
\caption{An example of advanced data visualization: Timeline and time-series plots 
	--- so-called stacked area plots, created in \textit{Scientific Python} (data analysis package 
	\textit{pandas}) mapping adaptation of Jupyter technology by authors of the paper during last three years (2017--2020).}
\label{fig:JupyterAdaptation}
\end{figure}

The second phase of our qualitative longitudinal study brought us another valuable information. 
Our three-year longitudinal research outlined the process of adapting Jupyter technology suitable for a teacher or a researcher in education. Using data science tools, we mapped the adaptation of Jupyter technology in our research and teaching practice via a longitudinal, time-series representation \citep{brockwell_introduction_2016} graphically summarized and visualized by Fig.~\ref{fig:JupyterAdaptation}. We analyzed all Jupyter notebooks,  created by all authors of the paper, by type, kernel and date using SciPy with pandas, a Python data analysis package originally developed in the context of financial time series modeling, with an extensive set of tools for working with time series data \citep{vanderplas_python_2016}.

The pandas stacked area plots in Fig.~\ref{fig:JupyterAdaptation} display the evolution of the daily average number notebooks created by authors (the overwhelming majority of Jupyter notebooks with R kernel was created by PS, with SageMath and Python kernels by JH) with respect to a timeline. The timeline shows the list of important events -- signing up to an online service or software installations. The first moment when we started to use Jupyter is connected to June 2017 when we signed up a CoCalc account (\href{https://cocalc.com}{cocalc.com}) to the last one, our submission of the paper (June 2020).

To avoid any problem with local installations mentioned by participants in the first phase, it seems as the best choice to start with Jupyter using one of the current cloud services with zero setup. We choose CoCalc (event: CoCalc account in Fig.~\ref{fig:JupyterAdaptation})\footnote{\footnotesize \textit{CoCalc} is a virtual online cloud workspace for calculations, research, collaboration and authoring documents combining the best free mathematical software and document editors. The service allows e.g. running Jupyter technology with many kernels (SageMath, Python, R, Julia etc.) and with real-time collaboration and communication tools directly in a web browser with zero setup. However, there are also another very similar and easy cloud ways to run Jupyter without any software installation \citep{data_school_six_2019}, e.g. you can use \textit{Binder} (\href{https://mybinder.org}{binder.org}) from Project Jupyter \citep{projectjupyter_binder_2018}, \textit{Kaggle kernels} (\href{https://www.kaggle.com/kernels}{kaggle.com/kernels}), \textit{Microsoft Azure notebooks} (\href{https://notebooks.azure.com}{notebooks.azure.com}) or \textit{Google Colab} (\href{https://colab.research.google.com}{colab.research.google.com}).}.
After a few months, we subscribed to a basic paid plan since the free trial running on free servers can be sometimes very slow (event: CoCalc subscription). 

With improving skills and better understanding we dared to try local installations
(Win-10 events). Being more familiar with the technology, we found that getting more benefits like more comfortable installation, work or better open-source packages means the transition to an open-source Linux operating system, in our case Fedora (event:Linux). Finally, our experience showed that leaving Windows is not an option for us. After two years we found a solution perfectly fitting requirements of effective research and teacher work -- the use of virtual machine Linux installation (via Oracle VM VirtualBox) which allow us to run Linux on Windows without shutting down Windows and combine the best features from the Windows and Linux world (VM events locally installing different Jupyter kernels: Sage, R, Julia, Octave, Fricas, PARI/GP).

\paragraph*{Discussion.} 

Results of our small qualitative case study suggest that a one-semester long course in using data science tools for Ph.D. PER candidates in the context of advanced statistical methods is not enough to start using them as main tools in their own research and analysis. Possible reasons indicated by the results include a short time (before our intervention for our students it was the first and only course in this field of expertise), steep learning curve and difficult reproducibility if we use the point-and-click environment. 

Students typically return to Excel as a natural and rescue option.  This appears in accordance with the general notion that Excel is probably still the most widespread digital tool for collecting, storing and processing educational data \citep{heiberger_r_2009,wilcox_understanding_2017}. 
However, young researchers use Excel, frequently with special add-ons allowing readily apply many advanced and modern methods, despite or unaware of the fact that overwhelming majority of their spreadsheets usually appear poorly or very hardly reproducible, with always present errors as it is in any point-and-click graphical interface depending on human factor\citep{panko_what_1998,baumer_modern_2017}\footnote{\footnotesize
We must say that R Commander has the build MarkDown system for writing easily reproducible and transparent reports. However, under the influence of guidebook \citep{heiberger_r_2009} we did not pay attention to this feature which became evident and very natural in Jupyter.}. This way of working with research data has been seen in wider circumstances and still belongs to not-negligible problems of current education-research publications \citep{van_der_zee_open_2018}. As for the shortness of learning time, our conclusion agrees with \cite{mckiernan_imagining_2017} who also pointed out that one semester for mastering skills with data tools is not enough.

Only after our intervention consisting in the exchange of learning tools, from the point-and-click environment (R Commander) to interactive multimedia documents (Jupyter Notebooks), and  the following continual advising, watching progress of our participants, we saw the full acceptance of data science tools which led also to creativity of participants. This result is connected with findings in \cite{odden_computational_2019,odden_physics_2019} where writing computational essays using Jupyter supports creative thinking.

As for the significance of participants' results, we can comment on the case in Fig.~\ref{fig:clusteplotscbig}. 
let us shortly explain the green cluster with higher concentration $C$ ($\geq 0.2$) and higher centered score $S$ ($\geq 0.25$). The higher $C$ means concentrating students' answers to less number of question choices (in density visualization it means peaks). The higher score S says that concentration is on the disagreement side with the statement. Therefore questions in the green cluster have approximately similar, higher concentration of students answers 'to one choice representing the disagreement. 

In PER the cluster analysis is still rare and perceived as difficult and ``magic''. EP was able to implement the hierarchical cluster analysis in her Ph.D. research, generally described in \cite{kassambara_practical_2017,emc_education_services_data_2015}, only after the Jupyter adaptation. We believe that the main reason why only several PER groups applied and reported this method up to now (e.g.\citet[SPSS]{ireson_multivariate_1999}, \citet[SAS]{ding_approaches_2009}, \citet[own C-code]{battaglia_unsupervised_2019}, \citet[Python code]{springuel_reconsidering_2019} consists in fact that all publications are missing the key components of open science \citep{van_der_zee_open_2018} -- analysis in an expensive commercial software or no available public data or no details of data analysis in a readable, easily reproducible code. 

To complete our comments on cluster analysis, 
Both ``convinced'' Ph.D. candidates strongly appreciated transparency and reproducibility of Jupyter notebooks consisting in the possibility to make own notes in along with computational procedures, analysis, plots, equations or algorithms. According to their experience, there was practically no problem to retrieve ideas developed in a notebook even a few months or a year after the end of work. Another highly valued feature was flexibility of open data science tools like R which is also generally accepted view
when they are compared with commercial softwares like SPSS \citep{wilcox_understanding_2017}.

Before getting the first knowledge about Jupyter in June 2017, from the mentioned EOSC project OpenDreamKit, all paper's authors, also teachers by their master degree (Math-Phys), had basic algorithmic skills and digital experience mainly with point-and-click statistical or mathematical environments. During three years, we successfully started with Jupyter, then naturally switched from the cloud to local work, from Windows to Linux and from two key kernels (SageMath, Python) to more kernels (R, Julia, Octave, etc.) -- the mode which is also typical and effective for the majority of current data scientists \citep{kaggle_kaggles_2020}. We apply Jupyter notebooks as an education tool and also as a scientific tool in our statistical, time-series research \citep{hancova_estimating_2020} and educational research (Ph.D. theses of EP and PS). Finally, it is important to say that observed time-series seasonal pattern (local maxima during summer months, minima during winter months) appears very specific and they are strongly determined by nature of university academic workflow and environment (July, August -- vacations; December, January -- exam periods).

\section*{Study 2: Pre-service physics teachers}

\vspace{6pt}
\subsection*{Background and context} 

The previous case study showed us valuable deep qualitative insights and detailed experience dealing with the use and perception of Jupyter technology from the viewpoint of ordinary educational researchers, from his first contact with the technology to its more advanced application. We were fully aware that type of the study, size of the sample and specific study circumstances would not allow valid generalizations. 

Therefore, using the results of the first phase of Study I, in the same year (2017), we grounded our approach in the following longitudinal quantitative study where the main goal was to explore teacher's perception of Jupyter in more standard conditions with a larger available sample of participants. 
In that case we decided to use the standardized, reliable quantitative diagnostic instrument called \textit{User Experience Questionnaire} \citep{laugwitz_construction_2008,schrepp_design_2017}, shortly UEQ. 

This marketing tool is widely used to measure a subjective, overall impression and satisfaction of user or ``customer'' experience with an interactive digital product. According to originators, the UEQ is the first tool covering user experience with the product satisfying  three important requirements -- a quick assessment; covering comprehensive impression; simple and immediate way to express affective attitudes. 

Particularly, the tool measures overall attractiveness of the product, its usability aspects (efficiency, perspicuity, dependability) and motivational aspects (stimulation, novelty). Perspicuity expresses how easy is to get familiar with the product and to learn how to use it. Efficiency stands for efficient and quick interaction with the product. Dependability expresses how the user feels in control of the interaction. Stimulation means motivation to use it and novelty measures how the innovativeness and creativeness of the product capture user's attention. 

\subsection*{Participants}
The participants were all future, pre-service teachers of physics, who always study physics in combination with other science or humanity subject, taking the first-year compulsory course of their bachelor study program, called \textit{Fundamentals of calculus for physicists} at the Institute of Physics at P.~J.~\v{S}af\'arik University in Ko\v{s}ice, Slovakia. We collected data after the course during three following years (2018-2020) from a total of $N = 33$ students (19 females, 14 males). Table~\ref{tab:PreTeachers} describes a numerical summary of participant demographics with respect to a year, gender and study field. The fourth factor, high or low achiever (Table~\ref{tab:Achievers}), was determined by a grade average \citep{nouri_flipped_2016}: high means an average from A to B, low represents an average from C to F.  

\begin{table}[!htpb]
\renewcommand{\arraystretch}{1.3}
\caption{Student demographics ($N=33$) in quantitative longitudinal study II.} \label{tab:PreTeachers}
\begin{tabular}{lccccccc}
	\hline
	&  &  \multicolumn{5}{c}{study field}& \\
	year & g &  Math-Phys &  Phys-Bio &  Phys-Chem &  Phys-Comp &  Phys-Geo   &  all   \\
	\hline
	\multirow{2}{*}{2017}  & F &          1 &         6 &          0 &          0 &         0 &    7 \\
	& M &          4 &         2 &          1 &          0 &         0 &    7 \\ \hline
	\multirow{2}{*}{2018} & F &          3 &         2 &          0 &          1 &         0 &    6 \\
	& M &          4 &         1 &          0 &          1 &         0 &    6 \\ \hline
	\multirow{2}{*}{2019}  & F &          3 &         2 &          0 &          0 &         1 &    6 \\
	& M &          1 &         0 &          0 &          0 &         0 &    1 \\ \hline
	& total &         16 &        13 &          1 &          2 &         1 &   33 \\
	\hline
\end{tabular}

\end{table}

\begin{table}[!htpb]
\renewcommand{\arraystretch}{1.3}
\caption{The longitudinal distribution of the sample as high or low achievers.} \label{tab:Achievers}
\begin{tabular}{lcccccccc}
	\hline
	study field & \multicolumn{2}{l}{Math-Phys} & \multicolumn{2}{l}{Phys-Bio} & Phys-Chem & Phys-Comp & Phys-Geo & all \\
	achiever &      high & low &     high & low &      high &       low & low \\
	\hline
	2017  &         4 &   1 &        2 &   6 &         1 &         0 &        0 &  14 \\
	2018  &         4 &   3 &        0 &   3 &         0 &         2 &        0 &  12 \\
	2019  &         3 &   1 &        1 &   1 &         0 &         0 &        1 &   7 \\ \hline
	total &        11 &   5 &        3 &  10 &         1 &         2 &        1 &  33 \\
	\hline
\end{tabular}
\end{table}

\subsection*{Design and methods}

\paragraph*{Instructional design of the course.}

The course, a bridge between secondary school and college level of mathematics and its study, focuses 
on the conceptual understanding, getting clear ideas and mastering basic computational and application skills connected with
fundamentals concepts of calculus: \textit{function, derivative, integral, differential equation} and \textit{complex number} in one and especially their more sophisticated versions in more dimensions. Our approach conceptually applies
ideas of the Calculus reform \citep{haver_calculus:_1999} which took place during the 80's and 90's of the previous century and completely re-thought the calculus curriculum for non-majors in math leading to such innovative college textbooks
like \citep{hughes-hallett_calculus:_2016}. 

After completing all essential and necessary preparation, from 2012 we teach the course in the frame of the Flipped learning \citep{talbert_flipped_2017, bergmann_flip_2012} using mentioned interactive teaching methods engaging student participation and active learning \citep{freeman_active_2014}. As for technology, we teach face-to-face, collaborative space group activities in a PC room whereas home preparation, individual space structured activities mainly based on interactive video lessons, are managed principally via Google Classroom as LMS and set of other supporting technologies, e.g. Edpuzzle (\href{https://edpuzzle.com/}{edpuzzle.com}) or smartphones.

An exploration, invention and application of mathematical concepts were realized by students with the help of two mathematical softwares Geogebra in the visual and graphical domain and Maxima in the analytical domain.  More details about our approach from content and pedagogical viewpoint can be found in \cite{hanc_application_2013,pankova_practical_2016,hanc_what_2016}.

Our major intervention made during June 2017-September 2017 consisted in replacing digital learning tools for active learning. Particularly, we transformed the majority of learning materials, Pdfs created in \LaTeX~typographical system,  directly to one uniform format --- online, interactive or static, Jupyter notebooks \citep{kueppers_latex_2017}.
The most important change was our transition from the point and click mathematical softwares Maxima to SageMath\footnote{\footnotesize SageMath is a free open-source Python-based mathematics software containing Maxima as its part  
\citep{stein_sage_2020,hogben_sage_2013,zimmermann_computational_2018} and now it is one of kernels in Jupyter. 
Originally SageMath was created as a free open source, viable alternative to the popular commercial computer algebra systems or scientific computing environments like Wolfram Mathematica (\href{https://www.wolfram.com/}{wolfram.com}), Maple (\href{https://www.maplesoft.com/}{maplesoft.com}), MATLAB (\href{https://www.mathworks.com/}{mathworks.com}).} as a Maxima replacement. 

\paragraph*{Data collection.} 
The basic available demographics and background information were obtained from our academic 
information system at the university. All personal information, respecting rules of GDPR, were 
deleted and basic identification was coded by random codes. 

Since the flipped math depends heavily on digital technologies and requires some basic level of
digital literacy from students, at the course beginning they filled out our self-developed \textit{Digital Experience Questionnaire} as an online Google form giving us contact information and their digital experience background. The questions are connected to own digital devices they can bring and use in collaborative activities during the face-to-face instruction, what in home preparation and what digital background in math education they bring from a secondary school (exact questions' wording is in the Appendix and our GitHub storage). 

At the end of the course, the mentioned UEQ tool was administrated, again electronically as an online Google form. Being recommended by UEQ originators, we translated the UEQ into Slovak\footnote{\footnotesize Up to these days, there are more than 30 language mutations of the original German version, including our Slovak version, all available at \href{https://www.ueq-online.org/}{ueq-online.org}.}. Its long version, used by us, contains 26 items grouped in six dimensions described above. In the UEQ, responses are in the form of the semantic differential where a respondent via a seven-point scale expresses to what extent he agrees with the given characteristic of the tool described by two opposite adjectives, e.g. \textit{annoying} $1\;\circlenumber{2}\;3\;4\;5\;6\;7$ \textit{enjoyable}.

To address the first personal viewpoint of how Jupyter contributed in the understanding of learning content, what self-progress and how transferable were own application of technology we extended UEQ by three 5-point Likert-scale questions (see the Appendix). The data collection was done three times with respect to the given conditions. 

\paragraph{Data analysis.} In statistical data analysis we applied the exploratory analysis based on basic descriptive numerical and advanced graphical summarizations allowed by data science tools (R and Python) and combined with standard ANOVA procedures \citep{maxwell_designing_2017} to test their statistical significance. The UEQ data has a special procedure for analysis described in \citep{laugwitz_construction_2008} with the following benchmark evaluation of a digital product for each dimension:
\vspace{6pt}
\begin{itemize}
\item  \textit{excellent}, your product is in the 10\% best product range,
\item  \textit{good}, 10\% of the best placed are higher and 75\% of the products  \\ are rated worse than your product,
\item  \textit{above average}, 25\% are better and 50\% worse than your product,
\item  \textit{below average}, 50\% are better and 25\% worse than your product,
\item  \textit{bad}, only 25\% of the products are worse than your product.
\end{itemize}
\vspace{6pt}
For UEQ data analysis we have also created our own R package containing functions suitable for our analysis. It can be found with all data, analysis and explanations in corresponding Jupyter notebooks at our GitHub repository \citep{hanc_jupyterper_2020-1}.

\subsection*{Results and discussion}

\paragraph{Initial Digital experience.} 

\begin{figure}[!h]
\includegraphics[width=0.97\linewidth]{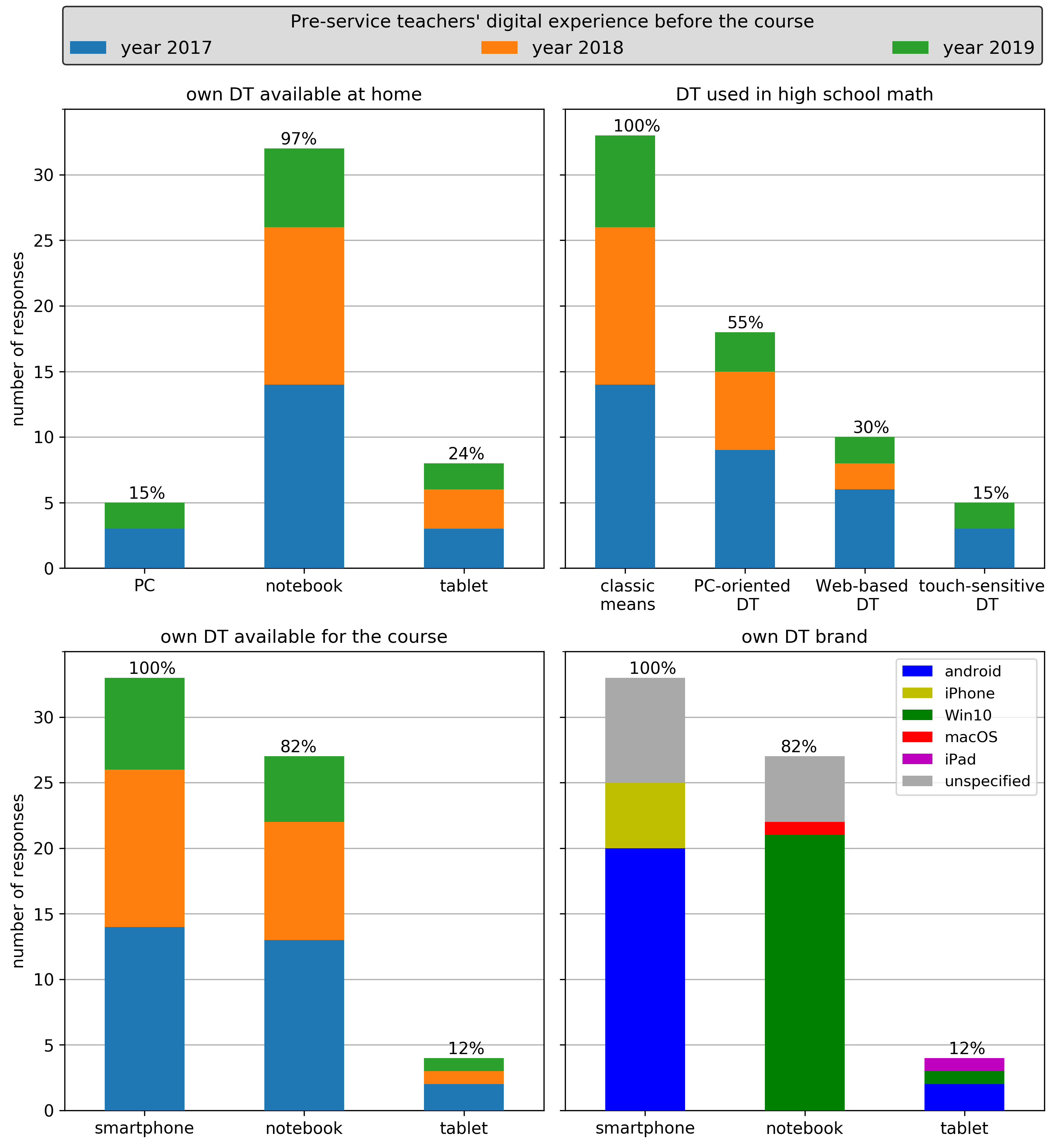}
\caption{\csentence{}
	Pre-service teacher's digital experience: own digital devices they can bring and use in the course, what in home preparation and what digital technology experienced in secondary school math education. Stacked bar plots were created in \textit{Scientific Python} (data analysis package 
	\textit{pandas}). }
\label{fig:digitalexperience}
\end{figure}

The longitudinal summary results of the Digital Experience Questionnaire data from 2017-2020 are depicted by Fig.~\ref{fig:digitalexperience}\footnote{\footnotesize Unlike other figures in the paper, the total percentages in the plots do not add up to 100 \% due to possible multiple student responses. Responses displayed by  each bar were considered independently with respect to the total possible number $N=33$.}. 
We can see that all students have access to several digital devices. Every student has their own smartphone (mainly Android). At home he can operate with a notebook (almost 100\%; 80\% have own one) or at least with a PC. The dominant operating system is Win10. A small part of students can use or bring a tablet. From the educational viewpoint, teaching and learning in secondary school math still rely on classical means (calculators, paper, chalk, blackboard). During a secondary school math instruction,  the half of students met PC-oriented technology (computers, notebooks, projectors, interactive whiteboard IWB), only  1/3 of them used web-based technology (LMS, email, videos or simulations) and a small number (15\%) touch-sensitive technology (smartphones and tablets).

\paragraph{User Experience Questionnaire.} From the original total $N=33$ students 15\% did not finish the first semester at their original bachelor study program. Specifically, 5 females: 2 (2017, low ach., Physics-Biology), 2 (2018, low ach., Physics-Biology, Physics-Computers), 1 (2018, high ach., Math-Physics) transferred to another study field (1 from Math-Physics to Math) or left our university (remaining 4). 

Therefore the UEQ data contains answers from $N^*=28$ students (14 females, 14 males).
Due to department politics, we always interview each student why he/she is leaving our study program and nobody saw our course as the key reason. Typical reasons were different expectations or too difficult introductory physics.

\begin{figure}[!h]
\includegraphics[width=0.97\linewidth]{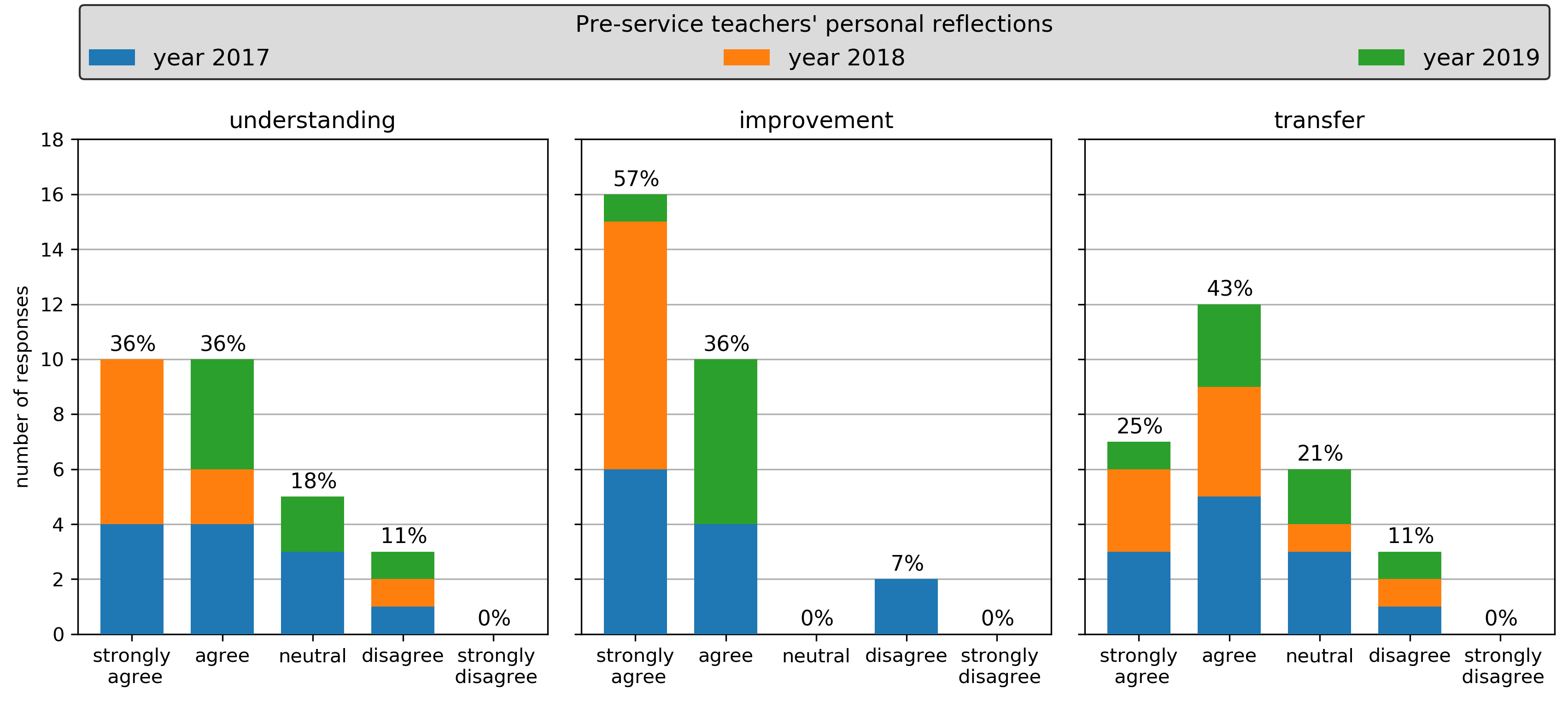}
\caption{\csentence{}
	Pre-service teacher's perception of technology role in understanding the learning content, in an improvement of own digital skills and in other subjects (possible transfer).}
\label{fig:TechnologyPerception}
\end{figure}

The first viewpoint of participants' perception represented by our three extra questions to UEQ (see graphical summary in Fig.~\ref{fig:TechnologyPerception}) show positive personal reflections of students  in the meaning of Jupyter for their learning. Explicitly, more than 70\% agreed that Jupyter Notebooks in our educational approach helped them a lot in understanding what they have learned. Overwhelming 90\% reported great improvement in the used digital technology.  Almost 80\% see a high potential of Jupyter use in other subjects. The longitudinal distributions of students' answers are qualitatively similar, although in understanding and improvement plots 2019-year modus is closer to the neutral level in comparison with previous years modi (2017, 2018).


\begin{figure}[]
\includegraphics[width=0.98\linewidth]{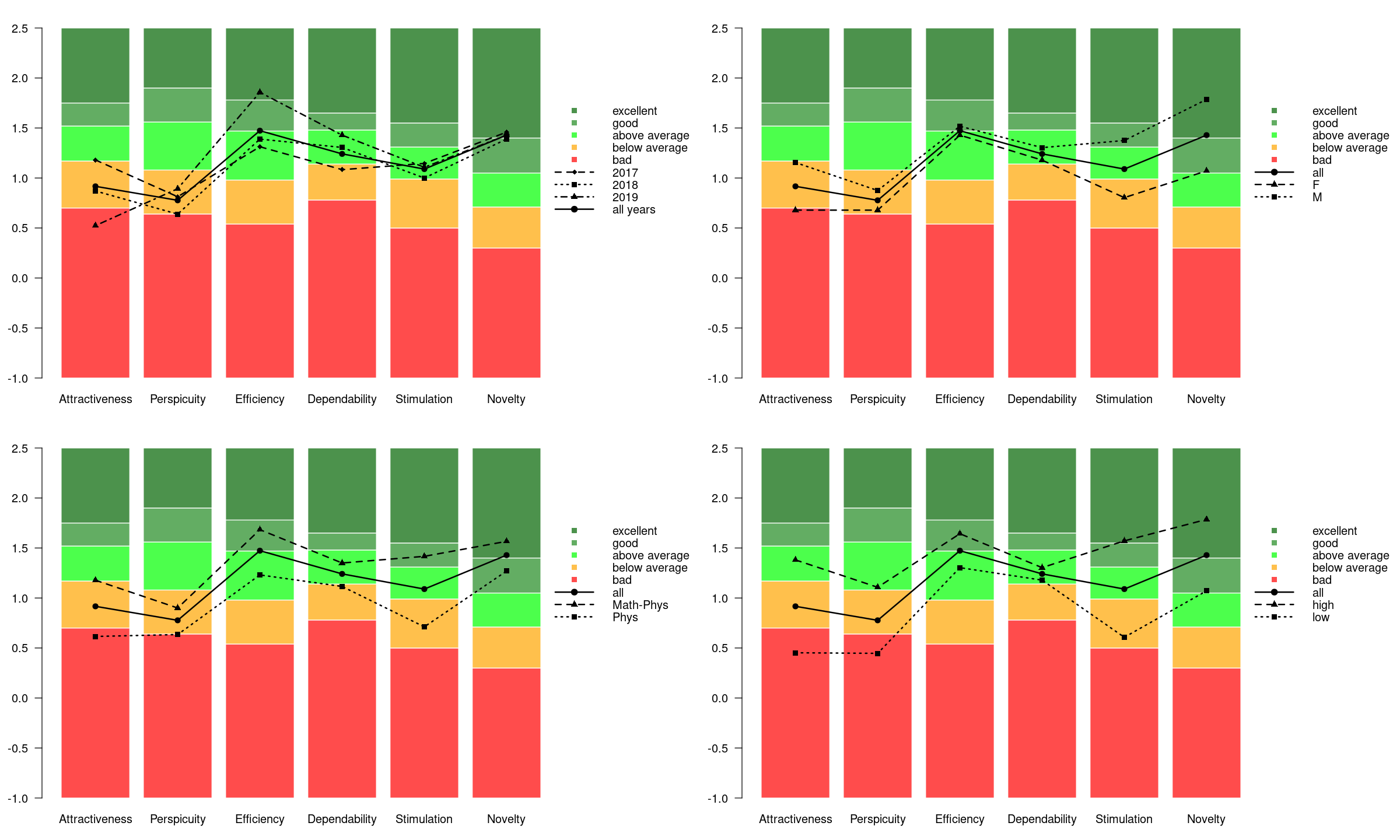}
\caption{\csentence{}
	The UEQ benchmark plots with respect to four factors -- year, gender, study program and 
	achievement. Plots were created in R (own UEQ package).}
\label{fig:UEQstudyII}
\end{figure}

The collected UEQ data, capturing students' impression, feelings and affective attitudes, led to the following benchmark results and interpretation summarized by Table~\ref{tab:UEQall}. The means of dimensions are also displayed in Fig.~\ref{fig:UEQstudyII} (solid line \textit{all}) which presents the UEQ benchmark results with respect to four factors -- year, gender, study program and achievement. 

\begin{table}[!h]
\renewcommand{\arraystretch}{1.3}
\caption{The UEQ benchmark results -- average (M), standard error (SE) and interpretation.} \label{tab:UEQall}
\begin{tabular}{lrll}
	\hline
	\multicolumn{1}{l}{dimension}&\multicolumn{1}{c}{M (SE)}&\multicolumn{1}{c}{benchmark}&\multicolumn{1}{c}{Interpretation}\tabularnewline
	\hline
	Attractiveness&0.92 (0.19)&below average&50\% better, 25\% worse\tabularnewline
	Perspicuity&0.78 (0.16)&below average&50\%  better, 25\%  worse\tabularnewline
	Efficiency&1.47 (0.17)&good&10\%  better, 75\%  worse\tabularnewline
	Dependability&1.24 (0.16)&above average&25\%  better, 50\%  worse\tabularnewline
	Stimulation&1.08 (0.20)&above average&25\%  better, 50\%  worse\tabularnewline
	Novelty&1.43 (0.15)&excellent&range of the best 10\%\tabularnewline
	\hline
\end{tabular}
\end{table}

These results demonstrate that no affective aspect of Jupyter is bad. Jupyter technology leaves the best personal impression in motivational aspect Novelty. The second most valuable for students is efficient and quick work with Jupyter. On the contrary,  Perspicuity, the usability aspect how easy to learn Jupyter, appears as a dimension with the weakest personal affective attitude. Very similar below average impression is connected to overall attractiveness of Jupyter as the digital product. 

Here it is worth to say that we aggregated all non-mathematical combinations to one level called shortly \textit{Phys}. This step is dictated not only by small numbers of students in Physics-Chemistry, Physics-Computers, Physics-Geography, but it is justified also conceptually since all non-math major programs have practically the same math and physics education, but different from the Math-Physics program with much wider and rigorous math education. 

Multiway ANOVA\footnote{\footnotesize As for the used inferential statistical analysis of the UEQ data, where each student evaluation is represented by six dimensional vector, MANOVA appears as a more suitable and powerful method. However we could not to apply it due to the insufficient size of our sample.} was used to test the significance of all four mentioned factors. As the graphical representation in Fig.~\ref{fig:UEQstudyII} indicates, ANOVA confirmed the statistical significance of factor ``achievement'' at Attractiveness ($p=0.020$, $\eta^2=0.192$ -- large effect size), Perspicuity ($p=0.048$, $\eta^2=0.142$ -- medium effect size), Stimulation ($p=0.043$, $\eta^2=0.135$ -- medium effect size) and Novelty ($p=0.017$, $\eta^2=0.159$ -- large effect size). Factor ``gender'' has a significant large effect at Novelty ($p=0.007$, $\eta^2=0.212$). 

The reliability of UEQ reached a high value -- Cronbach's $\alpha = 0.90 $ for the whole questionnaire and with respect to given dimensions alphas were between $ 0.55$ and $0.84$. The only dimension with questionable alpha was Dependability ($\alpha = 0.55$). All other values satisfy the generally accepted condition  $ \alpha> 0.7$ for reliable group measurements.

\paragraph*{Discussion.}
Our results, based on the questionnaires and our teaching observations, show that our future physics teachers as incoming secondary school students did not experience any digital technology as a key active learning tool in secondary school math education. Moreover, only a small proportion of them have required digital skills enhancing an effective and successful university study, especially in the mode of Flipped Learning. These findings explain why 90\% of students reported feelings of the great improvement in digital technology after our course.

Together with another implication from Study I, that one-semester long mathematical course with new technology can be refused by students, results also tell us that we have to be very careful in any digital innovation. Improper  instructional students' expectations and motivation with wrong meta-cognitive models about the role of learning tools (digital technology in our case) can be one of the decisive factors of low students' learning gains or failure \citep{hattie_visible_2009, hattie_applicability_2015}.

To avoid such situation and change students' expectations in the right direction as much as possible,  we realized several supporting steps from the beginning of our research which we believe became important determiners of our successful innovations. We prepared a new version of our supplementary supporting course called \textit{Students' digital literacy}. The original goal of the course was to provide the sufficient level of digital literacy in today's modern technologies (smartphone, tablet, social media via LMS, online web-technologies like Google drive) for better and more effective learning, active life in higher education and later professional activity. 

From 2017, regarding all university students -- future science teachers including given physics teachers, we incorporated in the course a special module for work and training with Jupyter and Geogebra. Since Jupyter does not have a special intuitive menu like point-and-click interfaces, one of the digital literacy goals also became developing very effective searching and working skills how to use practical cheat-sheets for Jupyter and its kernels which we collected and prepared in Slovak. 

Although creators and long-term Jupyter users feel Jupyter technology as simple as email or web \citep{projectjupyter_binder_2018}, our results suggest that there is a need to devote special attention to low achievers for whom Jupyter can be one of other unattractive obstacles in their study. Our interaction with such students led us to the finding that this is probably a problem of expectations. Low achievers have typically a negative attitude to math and automatically transmit it to everything which is connected with math. 

Concerning technological aspects, before 2017, our digital learning tools Maxima and Geogebra were treated by us and students as two independent tools. Now we have integrated SageMath with embedded Geogebra simulations and Jupyter widgets apps (in SageMath called Interacts) in Jupyter Notebooks as a unifying interactive multimedia environment\footnote{\footnotesize Technologically we run Jupyter with SageMath in three complementary modes: Binder for individual explorations, CoCalc for cooperative activities and SageMathCell (\href{https://sagecell.sagemath.org/}{sagecell.sagemath.org}) for very quick calculations. 
At home students usually install a Windows local version of the Jupyter with SageMath kernel.}. 
Interesting experience of our students showed that in the case of sufficiently large screens (at least 5 inches -- such smartphones are also called phablets) interactive Jupyter notebooks can be actively read and watch as study materials in home preparation. 

Finally, as for generalizations, this intervention single-case study without control group has again limits in external validity of results. However, our comparison of placement physics and math tests' results from our future physics teachers with results from secondary schools' population (see our third research study) suggest that our conclusion should represent upper limit what can be achieved. 

To complete our study report, we would like to mention that in our previous publication 
\citep{strauch_quantitative_2017} we showed high effectiveness of our flipped math course in the cognitive learning dimension, especially in the conceptual understanding of taught math concepts. From the content viewpoint,  the form of course was also principal. It provided form, tools and time for us and for students to get clear and transparent ideas about those math concepts which in such an early course with the traditional instructional model were considered as impossible to expose, understand or learn.

\section*{Study 3: In-service physics teachers}

\vspace{6pt}

\subsection*{Background and context}
The third parallel study, one of the main fresh, unpublished results of Ph.D. research of PS, deals with in-service physics teachers and their perception of the meaning of data and data science tools.
Generally, as for in-service teachers, there is a substantial body of research papers 
how quantitative and qualitative data can foster and improve effective teacher decisions and behavior \citep[see the special issue, Vol. 60, November 2016 in journal Teaching and Teacher 
Education, esp.][]{poortman_professional_2016, mandinach_what_2016, lai_impact_2016}. Simultaneously, 
data use belongs to the common and key characteristics of high-performing schools \citep{ebbeler_effects_2016}.

Data literacy for teaching, the ability to transform, collect, control and understand all types of educational data for better instruction, becomes essential. Every teacher should be also an action researcher armed at least by basic level of methodology knowledge and data literacy in his work for lifelong learning 
\citep{johnson_educational_2016}. Such action research will lead not only to a better understanding 
and improvement of own teaching practice, but above all to the greater satisfaction from work and a 
positive attitude to be better at their own profession.

Therefore the central goal of the Ph.D. thesis was to examine, verify and provide 
such data science tools, Jupyter as our first choice,  for in-service teachers in real practice which would help them easily and effectively to collect, process and use their own educational data. Specifically, we focused on data about students' preconceptions, misconceptions and mental models in physics understanding as one of the decisive factors in the quality of physics teaching \citep{fraser_teaching_2014, redish_oersted_2014}.

\subsection*{Participants}
\begin{figure}[!htpb]
\includegraphics[width=0.96\linewidth]{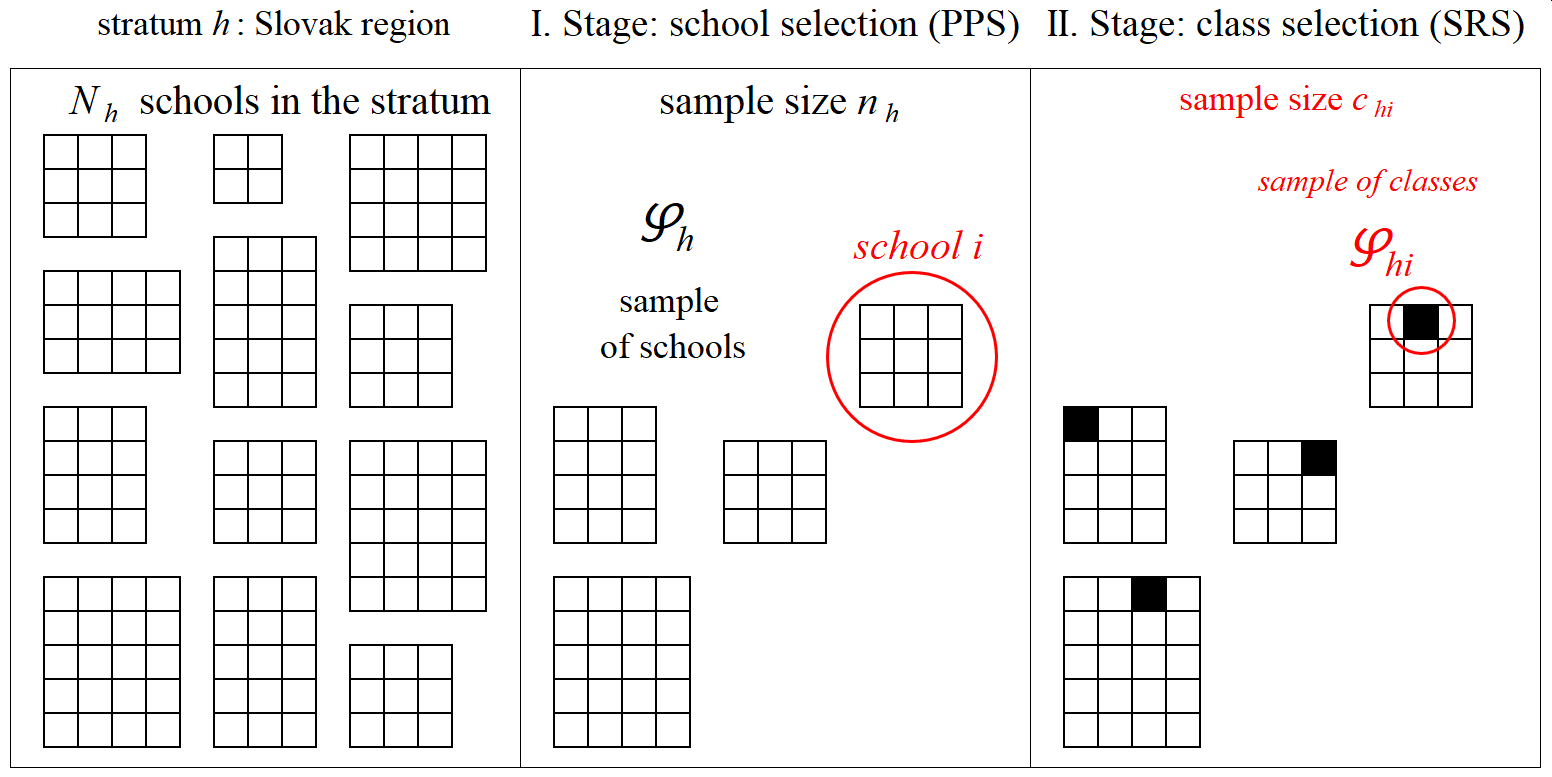}
\caption{A scheme of our complex survey, a stratified two-stage cluster sampling design, according to our applied version of the TIMSS (PISA) algorithm \citep[pp. 3.1-3.37]{laroche_sample_2016}. We took independently for each stratum a random probability proportional (PPS) sample of schools and simple random sampling (SRS) of classes at each chosen school. }
\label{fig:TIMSS}
\end{figure}

Participants were a random sample of $n = 40$ in-service physics secondary school teachers. 
Random sampling was a part of our national cross-sectional survey (May 2018-June 2018, 919 students) dealing with student conceptual understanding in mechanics with a stratified two-sample cluster sampling design according to an applied algorithm used in TIMSS (PISA) surveys (see a scheme in Fig.~\ref{fig:TIMSS}). 
Particularly, we took a random sample from all eligible Slovak grammar schools ($N = 284$), from $H =3 $ layers (strata), natural geographical Slovak regions -- western, central and eastern. From collected students' data we knew that all teachers successfully used our data science tool, however, online questionnaire feedback provided only $n^*=33$ teachers (83\% response rate). The left middle histogram in Fig.~\ref{fig:studyIII} in the result section shows demographic information about the sample of teachers (9 males and 24 females).

\subsection*{Design and methods}

\paragraph*{Pilot study.} We realized a pilot study in using Jupyter on a few small samples of our active physics teachers during physics teacher clubs. It showed negative feedback. Today, many requirements and complexity of teaching as a profession often lead to a very busy teacher, so in-service teachers are usually poorly concentrated on any longest adaptation of new technology also in the case of a simple one. Therefore, we left Jupyter and tried to find even a simpler data science tool.

\begin{figure}[!htpb]
\includegraphics[width=0.96\linewidth]{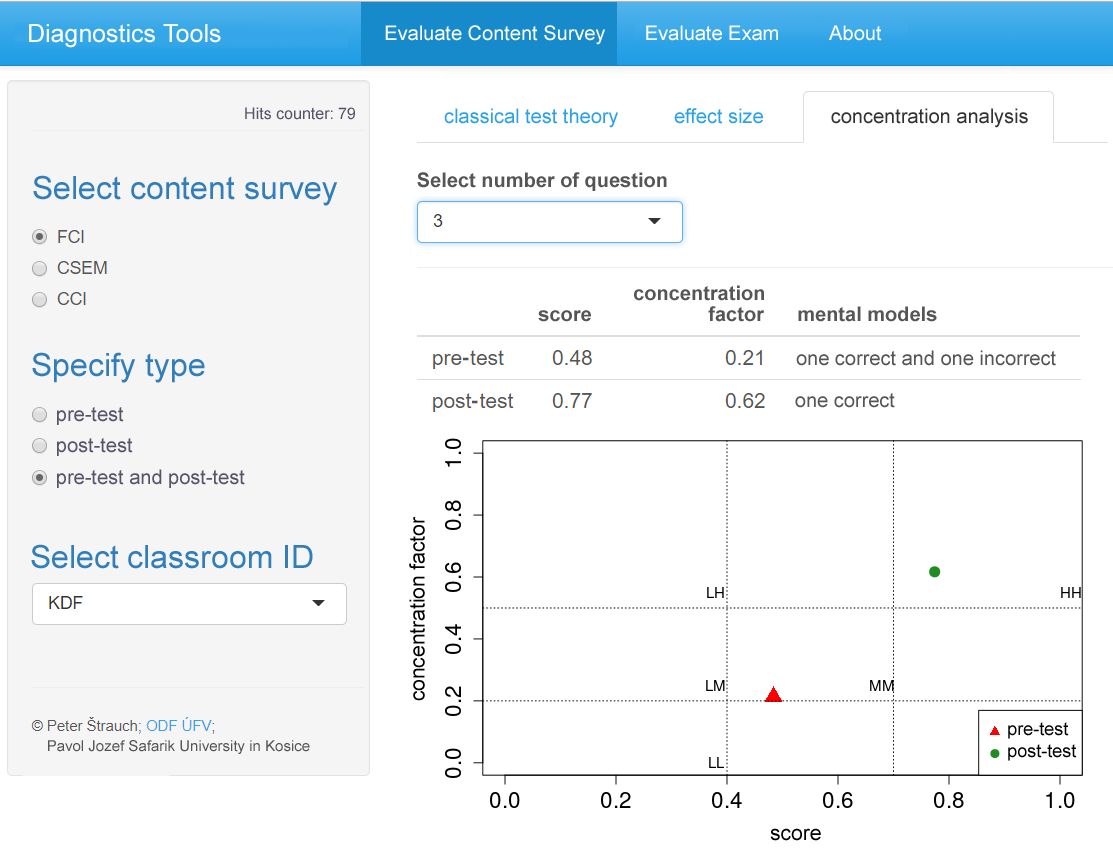}
\caption{A screenshot of our R shiny web app for open educational data analysis and visualization tested by  in-service physics teachers, randomly chosen from all eligible Slovak grammar schools.}
\label{fig:Shinyapp}
\end{figure}

\paragraph*{Data science tool: R Shiny.}
One of the revolutionary web  presentation tools of data science created in the R language is the R Shiny library \citep{chang_shiny:_2019}, which was released in 2012 by RStudio team (\href{https://shiny.rstudio.com/}{shiny.rstudio.com}). In open data and analysis, Shiny became the gold standard in interactive web presentation of data and statistics analyzes \footnote{\footnotesize
Shiny web apps have found application in many areas from statistics, through education, healthcare, genetics, pharmacy, geography, agriculture, or even economics, marketing or tourism, see e.g.
the official New Zealand tourism webpage  (\url{https://mbienz.shinyapps.io/tourism_dashboard_prod}).}.
Therefore using R Shiny library, we created own tailor-made web application which helps a teacher automatically collect his educational data from a Google form and transform it into interactive web presentation showing statistical analysis, graphical and numerical summaries with interpretation. A screenshot from this web app\footnote{\footnotesize The current version of our Shiny web app (now running only as Slovak after some extension) can be found at \url{https://odfufv.shinyapps.io/hodnotiaci-nastroj/}. It can be freely explored via a host code \textbf{CrbVUA}. Using Google translate, Google chrome extension, almost all parts of the web app are translated into English correctly.}, with a very intuitive point-and-click interface requiring no setup and zero adaptation learning time, is shown in Fig.~\ref{fig:Shinyapp}.

\paragraph{Data collection and analysis.} 
According to our instructions, during June 2018 teachers themselves own collected educational data from their physics classes via an online Google form offered by us in school PC rooms and subsequently on own device interactively ``played'' with data  in our R shiny app. After that teachers filled out our semi-closed self-developed attitude questionnaire (two items deal with the evaluation of R Shiny app, other ones are not directly relevant to this study, see the Appendix), and the UEQ evaluating our Shine web app. In statistical analysis we applied the same methods as in Study II whereas in UEQ analysis to get unbiased estimations, we incorporated calculations with weights according to the following general formula \citep{lohr_sampling_2009}:

\vspace{-12pt}

\begin{equation}
\bar{S} = \dfrac{\displaystyle\sum\limits_{h = 1}^{H} \;
\sum\limits_{i \in \mathscr{S}_{h}}
\sum\limits_{j \in \mathscr{S}_{h\nss i}}
\! w^*_{h \ns i} \cdot S_{h \ns i \ns j}}
{\displaystyle\sum\limits_{h =1}^{H} \;
\sum\limits_{i \in \mathscr{S}_{h \nss i}}
\sum\limits_{j \in \mathscr{S}_{h\nss i\nss j}}
\!\!	w^*_{h \ns i}
}, \qquad \displaystyle w^*_{h \ns i} \sim \dfrac{1}{n^*_h} \cdot \dfrac{U_h}{c_{h \ns i}^*} \sim 
\dfrac{1}{n^*_h} \cdot \dfrac{M_h}{c_{h \ns i}^*}
\end{equation}
where $S_{h \ns i \ns j}$ is a response $S$ (or its measure) from teacher $j$ from school $i$ in stratum $h$ ($h$ = 1, 2, 3) and $w_{hi}$ is the corresponding weight given by stratum properties: $n_h^*$ -- real number of participating teachers in the stratum, ${c_{h \ns i}^*}$ -- real number of participating classes from school $i$.  $U_h$ is the current total number of teachers and $M_h$ current total number of students in given stratum. All these numbers are available in data files at our GitHub storage.


\subsection*{Results and discussion}
\begin{figure}[h!]
\includegraphics[width=0.98\linewidth]{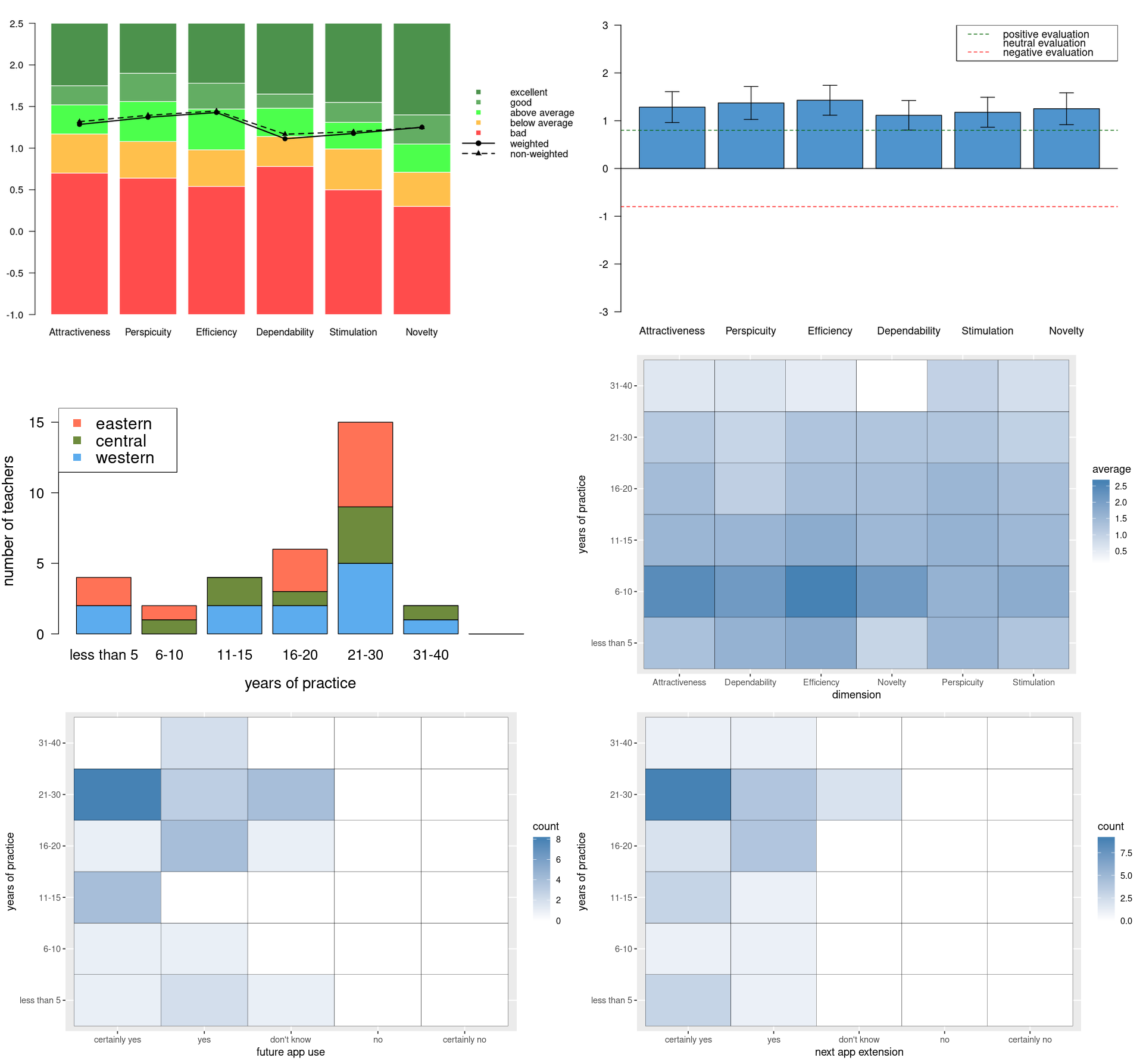}
\caption{\csentence{}
	Graphical visualization of teachers online feedback to our R shiny web application as a necessary replacement	of Jupyter. The left middle histogram shows demographic information about the sample of in-service teachers ($n^* = 33$, 9 males and 24 females).}
\label{fig:studyIII}
\end{figure}

The main results of our cross-sectional study connected to sample demographics, UEQ data for R Shiny app and attitude questionnaire data are summarized graphically in Fig.~\ref{fig:studyIII}. It contains a UEQ benchmark plot containing  original and weighted data (upper left plot), a bar chart with M and SE for all six UEQ dimensions (upper right plot). 
In the middle you can see demographics and a heat map of affective UEQ perceptions with respect to teacher's years of practice. The bottom left heat map displays teachers' attitude to use our web app in the future during
own instruction and the bottom right heat map teacher's attitude to the next extension of the web app. 


The UEQ benchmark shows that statistically correcting weights led to almost identical results -- differences between weighted and unweighted values ranged from 0.001 to 0.054. The most positive perception was expressed to Novelty (originality) of our digital product as good, which means that only 10\% of products are better, 75\% are worse. Attractiveness, Perspicuity and Stimulation were perceived and rated as above average (50\% other digital apps are worse and only 25\% have a  better impression). Only Dependability (confidence to work or control the product) with the weakest impression is on the border between above and below average. Cronbach's alphas were $0.96$ for the whole questionnaire and between $ 0.75$ and $0.90$ for all six dimensions, satisfying  condition  $ \alpha> 0.7$ for reliable group measurements.

Considering UEQ results from the viewpoint of the factor ``years of practice'', we got results (captured by the heat map in Fig.~\ref{fig:studyIII}) say that our web app created the best impression for teachers with 6-10 year practice for whom the app is excellent in all aspects. On the contrary, the app was least positively accepted by the most experienced teachers. 

As for willingness to use the app again in the feature, there was no negative attitude and more than 80\% want to use it again (see Tab.~\ref{tab:TAQ}). In the case of the future extension of our app to more tests and more subjects situation is even more positive and more than 90\%.
According to the corresponding heat map more positive attitudes are expressed by more experienced teachers.

\begin{table}[!h]
\renewcommand{\arraystretch}{1.3}
\caption{The teacher's attitudes to future use of our app and its extending.} \label{tab:TAQ}
\begin{tabular}{lrrr}
	\hline
	question &  strongly agree &  agree &  neutral  \\
	\hline
	future app use     &              45\% &     36\% &       18\%  \\
	next app extension &              58\% &     36\% &        6\%  \\
	\hline
\end{tabular}
\end{table}

\vspace{-12pt}

\paragraph*{Discussion.} 
Speaking about Jupyter failure in our case, later we realized that this finding is probably connected to the general implication \citep{mandinach_teachers_2016} that in the case of data literacy teachers must start with continuous lifelong learning as teacher candidates.  For teachers that are already in practice it is usually too late. Also similar experience leading to help physics teachers is presented in PhysPort recommendations \citep{mckagan_physport_2019,mckagan_physport:_2016}.

Very positive acceptation of R Shiny web app among our in-service teachers is really promising. Since Shiny app provide also overall summaries from all teachers using the application, a teacher can compare results of his students with others which appears to be one of the big motivational benefits. Considering R Shiny not only as an open data tool, but also as a learning tool, we noticed many papers reporting a positive experience and educational potential of its application \cite[e.g.][]{fawcett_using_2018, doi_web_2016}. However, open attitude questionnaire comments of teachers and our interaction with teachers signalize problems in the clear and deeper conceptual understanding of data visualization and its statistical interpretation (statistical literacy overlapping with data literacy, \cite{gould_data_2017}). 

\section*{Conclusions}

\vspace{6pt}

\subsection*{Key findings}
Open source digital data science tools over the past 10 years, especially Jupyter and R, have provided us with revolutionary, free, highly accessible ways to collect, store, process, or share data. In light of these changes, not only science, economics but also education is changing. 

Our theoretical analysis and research pointed out that there is a vast body of publications dealing with Jupyter application as a web-based interactive educational digital tool in different types of instruction intervention. Our pedagogical review also demonstrates the unifying nature of Jupyter technology. Jupyter notebooks interactively combine all forms of information together as one multimedia tool. Jupyter can also combine different educational tools under one hood providing important means for current modern interactive teaching approaches (PI, QDI, IBSE, FL), e.g. immediate feedback via e-voting (Jupyter activity extension), virtual active experimentation (Jupyter widget and embedding functionality) or interactive video-lessons via screencasting (Jupyter Graffiti extension).

Empirical educational research dealing with Jupyter implementation only recently started to apply general principles of the cognitive psychological theory of multimedia and online learning which themselves need the stronger incorporation of the affective and motivational dimension of learning. 

In this paper we presented our three empirical studies combined in the convergent design of the three-years long complex mixed-method research whose goal was to examine affective aspects and perception of Jupyter implementation during higher and after higher (lifelong) education, in three groups of physics teachers: PER candidates and research, future pre-service teachers and current in-service teachers.

Using self-developed qualitative and self-developed and standardized quantitative diagnostic tools, we uncovered the findings not only consistent with other, closely connected studies, but also highlighting specific, additional implications resulting from Jupyter application as a learning tool. 
Here we like to draw attention to the standardized, reliable quantitative diagnostic instrument called User Experience Questionnaire (UEQ) \citep{laugwitz_construction_2008} which is widely used to measure a subjective, overall impression and satisfaction of user, “customer” experience with any digital product.
Our contribution is own, free available R implementation of UEQ analysis extended by the possibility to use weighted data.

Our results of Study I and II indicate that young  beginning PER researchers and pre-service physics teachers can master key digital skills for work with Jupyter technology, positively appreciating its big impact on their learning, data and statistical literacy or professional development.
These results support the ongoing worldwide effort to implement Jupyter in
traditional education as a promising free open-source interactive learning tools to
foster learning process, especially for young generation.

Despite the fact that open-source Jupyter notebooks are natural and easy as email or web, results of Study III suggest that in-service teachers are not prepared for Jupyter technology as a tool supporting their more effective work and instructional steps. This finding is in accordance with the general conclusion dealing with getting data literacy for teaching. However, we found that they can work and very positively accept open education data presented via another open-source data science tool, self-developed R Shiny interactive web application, as an important form of immediate feedback and learning about the quality of their instruction. Younger teachers appreciate affective aspects of R Shiny web app and older immediate data feedback.

\subsection*{Implications for practice and further research}

Here we highlight several important findings to avoid failures. Overcoming possible barriers in adapting data science tools like Jupyter with the appropriate kernel (R or Python) consists in using today free cloud services with zero setup at the beginning and apply just-in-time local installations of technology, ending with the mode of workflow typical for a current data scientist. 
Concerning technological installation aspects, a very effective and time-saving Jupyter implementation is in a virtual machine Linux running on Win10. 
Moreover, it seems that Jupyter features like easy to share, reproducibility, transparency and flexibility can also help a beginner to master advanced research methods of data collection, processing, analysis and visualization, e.g. time series visualization or cluster analysis. By graphical and statistical results our paper demonstrates what can be done in this direction. 

From the educational viewpoint, a one-time or short adaptation of Jupyter can be unsuccessful. We recommend to pay special attention to low achievers and females who under wrong expectations can perceive Jupyter technology as more difficult than in reality is. Such special attention can be
realized e.g. via an additional digital literacy course supporting effective learning with Jupyter and changing wrong expectations in the right direction. 

In connection with R Shiny and the mentioned European open science cloud service, we see not only educational researchers, university students, but also every modern teacher as an active and effective 
participant in using open data, technologies and e-infrastructure mediated by the EOSC on daily 
basis. Speaking about data and digital literacy skills of teachers and the data meaning for effective teaching \citep{mandinach_teachers_2016, 	mandinach_what_2016, sergis_using_2019,ndukwe_teaching_2020,van_den_hurk_fostering_2016},
EOSC will allow to process their own educational data or find one from all publicly available research in 
Europe in a few clicks, combine them in new, sophisticated ways without a significant loss of time, 
without expensive or special hardware, operating in any modern Internet browser without 
the need for installation or complex administration.

Finally it is necessary to say that results of our research (all three studies were intervention single-cased or observational) cannot be fully generalized and must be confirmed or disproved by future wider experimental quantitative research. On other hand, we have to say that educational research in this field is still in its infancy so there is still an urgent need to collect and summarize any available knowledge and experience which can be later important in realizing more valid and  general experimental studies.



\begin{backmatter}

\section*{Abbreviations}
STEM: science, technology, engineering and mathematics; STEAM: science, technology, engineering, arts and mathematics; MOOC: massive open online courses; IBSE: inquiry based science education; PI: peer instruction; QDI: question driven instruction; FL: flipped learning;  PER: physics education research; SPSS: Statistical Package for the Social Sciences; RSRP: Real Statistics Resource Pack; AAPT: The American Association of Physics Teachers; VM: virtual machine software; UEQ: User Experience Questionnaire; PPS: random probability proportional sampling; SRS: simple random sampling.


\section*{Acknowledgments}
The authors would like to thank Andrej Gajdo\v{s} for his advice in using R in advanced multivariate statistical analysis. In connection with applications of cluster analysis in physics education research our special thanks go to R. Padraic Springuel and Onofrio R. Battaglia for communication dealing with their own cluster analysis code.

\section*{Funding}
This work was supported by the Slovak Research and Development Agency under the contract No.
APVV-17-0568.



\bibliographystyle{spbasic} 
\bibliography{references}      

\begin{thebibliography}{121}
\providecommand{\natexlab}[1]{#1}
\providecommand{\url}[1]{{#1}}
\providecommand{\urlprefix}{URL }
\expandafter\ifx\csname urlstyle\endcsname\relax
  \providecommand{\doi}[1]{DOI~\discretionary{}{}{}#1}\else
  \providecommand{\doi}{DOI~\discretionary{}{}{}\begingroup
  \urlstyle{rm}\Url}\fi
\providecommand{\eprint}[2][]{\url{#2}}

\bibitem[{Addinsoft(2020)}]{addinsoft_xlstat_2020}
Addinsoft (2020) {XLSTAT} {\textbar} {Statistical} {Software} for {Excel}.
  \urlprefix\url{https://www.xlstat.com/en/}

\bibitem[{Angerer et~al(2020)Angerer, Kluyver, Schulz, abielr, Sa, Hester,
  karldw, Foster, and Sievert}]{angerer_repr_2020}
Angerer P, Kluyver T, Schulz J, abielr, Sa DFd, Hester J, karldw, Foster D,
  Sievert C (2020) repr: {Serializable} {Representations}.
  \urlprefix\url{https://CRAN.R-project.org/package=repr}

\bibitem[{Bao and Redish(2001)}]{bao_concentration_2001}
Bao L, Redish EF (2001) Concentration analysis: {A} quantitative assessment of
  student states. Am J Phys 69(7):45--53, \doi{10.1119/1.1371253}

\bibitem[{Barba et~al(2019)Barba, Barker, Blank, Brown, Downey, George, Heagy,
  Mandli, Moore, Lippert, Niemeyer, Watkins, West, Wickes, Willing, and
  Zingale}]{barba_teaching_2019}
Barba LA, Barker LJ, Blank DS, Brown J, Downey AB, George T, Heagy LJ, Mandli
  KT, Moore JK, Lippert D, Niemeyer KE, Watkins RR, West RH, Wickes E, Willing
  C, Zingale M (2019) Teaching and {Learning} with {Jupyter}. GitHub, Creative
  Commons Attribution CC-BY 4.0 International license,
  \urlprefix\url{https://jupyter4edu.github.io/jupyter-edu-book/}

\bibitem[{Battaglia et~al(2019)Battaglia, Di~Paola, and
  Fazio}]{battaglia_unsupervised_2019}
Battaglia OR, Di~Paola B, Fazio C (2019) Unsupervised quantitative methods to
  analyze student reasoning lines: {Theoretical} aspects and examples. Phys Rev
  Phys Educ Res 15(2):020,112, \doi{10.1103/PhysRevPhysEducRes.15.020112}

\bibitem[{Baumer et~al(2017)Baumer, Kaplan, and Horton}]{baumer_modern_2017}
Baumer BS, Kaplan DT, Horton NJ (2017) Modern {Data} {Science} with {R}, 1st
  edn. Chapman and Hall/CRC, Boca Raton

\bibitem[{Beatty and Gerace(2009)}]{beatty_technology-enhanced_2009}
Beatty ID, Gerace WJ (2009) Technology-{Enhanced} {Formative} {Assessment}: {A}
  {Research}-{Based} {Pedagogy} for {Teaching} {Science} with {Classroom}
  {Response} {Technology}. Journal of Science Education and Technology
  18(2):146--162, \doi{10.1007/s10956-008-9140-4}

\bibitem[{Beatty et~al(2006)Beatty, Gerace, Leonard, and
  Dufresne}]{banks_question_2006}
Beatty ID, Gerace WJ, Leonard WJ, Dufresne RJ (2006) Question driven
  instruction: {Teaching} science (well) with an audience response system,
  chap. 7. In: Banks DA (ed) Audience {Response} {Systems} in {Higher}
  {Education}: {Applications} and {Cases}, Information Science Publishing,
  London, pp 96--115

\bibitem[{Beezer et~al(2013)Beezer, Bradshaw, Grout, and
  Stein}]{hogben_sage_2013}
Beezer RA, Bradshaw R, Grout J, Stein WA (2013) Sage. In: Hogben L (ed)
  Handbook of {Linear} {Algebra}, 2nd edn, Chapman and Hall/CRC, Boca Raton, pp
  91--1--91--26

\bibitem[{Bergmann and Sams(2012)}]{bergmann_flip_2012}
Bergmann J, Sams A (2012) Flip {Your} {Classroom}: {Reach} {Every} {Student} in
  {Every} {Class} {Every} {Day}. International Society for Technology in
  Education, New York

\bibitem[{Blank and Silvester(2020)}]{blank_calystometakernel_2020}
Blank DS, Silvester S (2020) Calysto/metakernel.
  \urlprefix\url{https://github.com/Calysto/metakernel}, original-date:
  2014-08-24T20:53:45Z

\bibitem[{Boaler(2020)}]{boaler_data_2020}
Boaler J (2020) Data {Science} {Initiative} {Video}.
  \urlprefix\url{https://www.youcubed.org/resources/data-science-initiative-video/}

\bibitem[{Boisvert et~al(2016)Boisvert, Born, Bowen, Gould, Kapaun, Konold,
  Lavista~Ferres, Merchant, Shaffner, Whitney, and
  Williams}]{boisvert_building_2016}
Boisvert D, Born K, Bowen M, Gould R, Kapaun R, Konold C, Lavista~Ferres JM,
  Merchant O, Shaffner A, Whitney H, Williams M (2016) Building {Global}
  {Interest} in {Data} {Literacy}: {A} {Dialogue}. Workshop report, Oceans of
  Data Institute, Education Development Center, Inc., Waltham,
  \urlprefix\url{http://oceansofdata.org/our-work/building-global-interest-data-literacy-dialogue-workshop-report}

\bibitem[{Brockwell and Davis(2016)}]{brockwell_introduction_2016}
Brockwell PJ, Davis RA (2016) Introduction to {Time} {Series} and
  {Forecasting}, 3rd edn. Springer, New York

\bibitem[{Bu and Schoen(2011)}]{bu_model-centered_2011}
Bu L, Schoen R (eds)  (2011) Model-centered {Learning}: {Pathways} to
  {Mathematical} {Understanding} {Using} {GeoGebra}. Springer, New York

\bibitem[{Cardoso et~al(2019)Cardoso, Leit{\~a}o, and
  Teixeira}]{cardoso_using_2019}
Cardoso A, Leit{\~a}o J, Teixeira C (2019) Using the {Jupyter} {Notebook} as a
  {Tool} to {Support} the {Teaching} and {Learning} {Processes} in
  {Engineering} {Courses}. In: Auer ME, Tsiatsos T (eds) The {Challenges} of
  the {Digital} {Transformation} in {Education}, Springer International
  Publishing, Cham, Advances in {Intelligent} {Systems} and {Computing}, pp
  227--236, \doi{10.1007/978-3-030-11935-5_22}

\bibitem[{Chang et~al(2019)Chang, Cheng, Allaire, Xie, McPherson, RStudio,
  library), inst/www/shared/jquery AUTHORS.txt),
  inst/www/shared/jqueryui/AUTHORS.txt), library), library), library), Twitter,
  library), library), library), library), library), font), library), library),
  library), library), library), library), library), library), library), and
  R)}]{chang_shiny:_2019}
Chang W, Cheng J, Allaire JJ, Xie Y, McPherson J, RStudio, library) jFjlajU,
  inst/www/shared/jquery AUTHORStxt) jcjlali,
  inst/www/shared/jqueryui/AUTHORStxt) jUcjUlali, library) MOB, library) JTB,
  library) BcB, Twitter, library) IB, library) AFh, library) SJRj, library)
  SPBd, library) ARBd, font) DGFA, library) BRsj, library) KMKes, library)
  esces, library) DIir, library) SSJs, library) SLD, library) JFsj, library)
  JGsj, library) IShj, R) RCTtif (2019) shiny: {Web} {Application} {Framework}
  for {R}. \urlprefix\url{https://CRAN.R-project.org/package=shiny}

\bibitem[{Constantinou et~al(2018)Constantinou, Tsivitanidou, and
  Rybska}]{constantinou_what_2018}
Constantinou CP, Tsivitanidou OE, Rybska E (2018) What {Is} {Inquiry}-{Based}
  {Science} {Teaching} and {Learning}? In: Tsivitanidou OE, Gray P, Rybska E,
  Louca L, Constantinou CP (eds) Professional {Development} for
  {Inquiry}-{Based} {Science} {Teaching} and {Learning}, Contributions from
  {Science} {Education} {Research}, Springer International Publishing, Cham, pp
  1--23, \doi{10.1007/978-3-319-91406-0_1}

\bibitem[{Creswell and Clark(2017)}]{creswell_designing_2017}
Creswell JW, Clark VLP (2017) Designing and {Conducting} {Mixed} {Methods}
  {Research}, 3rd edn. SAGE Publications, Inc, London

\bibitem[{Crouch and Mazur(2001)}]{crouch_peer_2001}
Crouch CH, Mazur E (2001) Peer {Instruction}: {Ten} years of experience and
  results. American Journal of Physics 69(9):970--977, \doi{10.1119/1.1374249}

\bibitem[{{Data School}(2019)}]{data_school_six_2019}
{Data School} (2019) Six easy ways to run your {Jupyter} {Notebook} in the
  cloud.
  \urlprefix\url{https://www.dataschool.io/cloud-services-for-jupyter-notebook/}

\bibitem[{Ding and Beichner(2009)}]{ding_approaches_2009}
Ding L, Beichner R (2009) Approaches to data analysis of multiple-choice
  questions. Phys Rev ST Phys Educ Res 5(2):020,103.1--020,103.17,
  \doi{10.1103/PhysRevSTPER.5.020103}

\bibitem[{Doi et~al(2016)Doi, Potter, Wong, Alcaraz, and Chi}]{doi_web_2016}
Doi J, Potter G, Wong J, Alcaraz I, Chi P (2016) Web {Application} {Teaching}
  {Tools} for {Statistics} {Using} {R} and {Shiny}. Technology Innovations in
  Statistics Education 9(1),
  \urlprefix\url{https://escholarship.org/uc/item/00d4q8cp}

\bibitem[{Downes(2019)}]{downes_look_2019}
Downes S (2019) A {Look} at the {Future} of {Open} {Educational} {Resources}.
  International Journal of Open Educational Resources 1(2):33--49,
  \doi{10.18278/ijoer.1.2.4}

\bibitem[{Ebbeler et~al(2016)Ebbeler, Poortman, Schildkamp, and
  Pieters}]{ebbeler_effects_2016}
Ebbeler J, Poortman CL, Schildkamp K, Pieters JM (2016) Effects of a data use
  intervention on educators{\textquoteright} use of knowledge and skills.
  Studies in Educational Evaluation 48:19--31,
  \doi{10.1016/j.stueduc.2015.11.002}

\bibitem[{{EDC Oceans of Data
  Institute}(2015)}]{edc_oceans_of_data_institute_building_2015}
{EDC Oceans of Data Institute} (2015) Building {Global} {Interest} in {Data}
  {Literacy}: {A} {Dialogue} {\textbar} {Oceans} of {Data}.
  \urlprefix\url{http://oceansofdata.org/projects/building-global-interest-data-literacy-dialogue}

\bibitem[{{EMC Education Services}(2015)}]{emc_education_services_data_2015}
{EMC Education Services} (2015) Data {Science} and {Big} {Data} {Analytics}:
  {Discovering}, {Analyzing}, {Visualizing} and {Presenting} {Data}, 1st edn.
  Wiley, Indianapolis

\bibitem[{Fawcett(2018)}]{fawcett_using_2018}
Fawcett L (2018) Using {Interactive} {Shiny} {Applications} to {Facilitate}
  {Research}-{Informed} {Learning} and {Teaching}. Journal of Statistics
  Education 26(1):2--16

\bibitem[{Fletcher(2010)}]{fletcher_psychometric_2010}
Fletcher TD (2010) psychometric: {Applied} {Psychometric} {Theory}.
  \urlprefix\url{https://CRAN.R-project.org/package=psychometric}

\bibitem[{Fox(2016)}]{fox_using_2016}
Fox J (2016) Using the {R} {Commander}: {A} {Point}-and-{Click} {Interface} for
  {R}, 1st edn. Chapman and Hall/CRC, Milton

\bibitem[{Fox et~al(2020{\natexlab{a}})Fox, Bouchet-Valat, Andronic, Ash, Boye,
  Calza, Chang, Grosjean, Heiberger, Pour, Kerns, Lancelot, Lesnoff, Ligges,
  Messad, Maechler, Muenchen, Murdoch, Neuwirth, Putler, Ripley, Ristic, Wolf,
  and Wright}]{fox_rcmdr_2020}
Fox J, Bouchet-Valat M, Andronic L, Ash M, Boye T, Calza S, Chang A, Grosjean
  P, Heiberger R, Pour KK, Kerns GJ, Lancelot R, Lesnoff M, Ligges U, Messad S,
  Maechler M, Muenchen R, Murdoch D, Neuwirth E, Putler D, Ripley B, Ristic M,
  Wolf P, Wright K (2020{\natexlab{a}}) Rcmdr: {R} {Commander}.
  \urlprefix\url{https://CRAN.R-project.org/package=Rcmdr}

\bibitem[{Fox et~al(2020{\natexlab{b}})Fox, Muenchen, and
  Putler}]{fox_rcmdrmisc_2020}
Fox J, Muenchen R, Putler D (2020{\natexlab{b}}) {RcmdrMisc}: {R} {Commander}
  {Miscellaneous} {Functions}.
  \urlprefix\url{https://CRAN.R-project.org/package=RcmdrMisc}

\bibitem[{Fraser et~al(2014)Fraser, Timan, Miller, Dowd, Tucker, and
  Mazur}]{fraser_teaching_2014}
Fraser JM, Timan AL, Miller K, Dowd JE, Tucker L, Mazur E (2014) Teaching and
  physics education research: bridging the gap. Reports on Progress in Physics
  77(3):032,401, \doi{10.1088/0034-4885/77/3/032401}

\bibitem[{Frederickson(2019)}]{frederickson_ranking_2019}
Frederickson B (2019) Ranking {Programming} {Languages} by {GitHub} {Users}.
  \urlprefix\url{https://www.benfrederickson.com/ranking-programming-languages-by-github-users/}

\bibitem[{Freeman et~al(2014)Freeman, Eddy, McDonough, Smith, Okoroafor, Jordt,
  and Wenderoth}]{freeman_active_2014}
Freeman S, Eddy SL, McDonough M, Smith MK, Okoroafor N, Jordt H, Wenderoth MP
  (2014) Active learning increases student performance in science, engineering,
  and mathematics. PNAS 111(23):8410--8415, \doi{10.1073/pnas.1319030111}

\bibitem[{{GitHub, Inc.}(2020)}]{github_inc_github_2020}
{GitHub, Inc} (2020) {GitHub} {Milestones}.
  \urlprefix\url{https://github.com/about/milestones}

\bibitem[{Gould(2017)}]{gould_data_2017}
Gould R (2017) Data {Literacy} is {Statistical} {Literacy}. Statistics
  Education Research Journal 16(1):22--25

\bibitem[{Hake(1998)}]{hake_interactive-engagement_1998}
Hake RR (1998) Interactive-engagement versus traditional methods: {A}
  six-thousand-student survey of mechanics test data for introductory physics
  courses. American Journal of Physics 66(1):64--74, \doi{10.1119/1.18809}

\bibitem[{Hall and Lingefj{\"a}rd(2016)}]{hall_mathematical_2016}
Hall J, Lingefj{\"a}rd T (2016) Mathematical {Modeling}: {Applications} with
  {GeoGebra}. John Wiley \& Sons, Hoboken

\bibitem[{Han{\v c}(2013)}]{hanc_application_2013}
Han{\v c} J (2013) Application of the flipped classroom model in science and
  math education in {Slovakia}. In: {HSCI} 2013: {Proceedings} of the 10th
  {International} conference on {Hands}-on {Science} (1st-5th {July} 2013,
  {Ko{\v s}ice}), P.J. {\v S}af{\'a}rik University, Ko{\v s}ice, Slovakia, pp
  229--234

\bibitem[{Han{\v c}(2016)}]{hanc_what_2016}
Han{\v c} J (2016) What is going on in {Slovakia}? {Current} trends and flipped
  learning. In: Santiago R (ed) Actas del {II} {Congreso} de {Flipped}
  {Classroom}: {Comunicaciones} y posters presentados, Edita MT Servicios
  Educativos, Zaragoza, Spain, pp 328--344

\bibitem[{Han{\v c} et~al(2011)Han{\v c}, Luk{\'a}{\v c}, Seker{\'a}k, and {\v
  S}veda}]{hanc_geogebra_2011}
Han{\v c} J, Luk{\'a}{\v c} S, Seker{\'a}k J, {\v S}veda D (2011) Geogebra
  {\textemdash} {A} complex digital tool for highly effective math and science
  teaching. In: 2011 9th {International} {Conference} on {Emerging} {eLearning}
  {Technologies} and {Applications} ({ICETA}), Elfa, Ko{\v s}ice, pp 131--136,
  \doi{10.1109/ICETA.2011.6112601}

\bibitem[{Han{\v c} et~al(2020{\natexlab{a}})Han{\v c}, {\v S}trauch, and
  Han{\v c}ov{\'a}}]{hanc_jupyterper_2020}
Han{\v c} J, {\v S}trauch P, Han{\v c}ov{\'a} M (2020{\natexlab{a}})
  {JupyterPer}. \urlprefix\url{https://github.com/JupyterPER}

\bibitem[{Han{\v c} et~al(2020{\natexlab{b}})Han{\v c}, {\v S}trauch, and
  Han{\v c}ov{\'a}}]{hanc_jupyterper_2020-1}
Han{\v c} J, {\v S}trauch P, Han{\v c}ov{\'a} M (2020{\natexlab{b}})
  {JupyterPER}: {Open}-{Education}-{Science}.
  \urlprefix\url{https://github.com/JupyterPER/Open-Education-Science}

\bibitem[{Han{\v c}ov{\'a} et~al(2020)Han{\v c}ov{\'a}, Gajdo{\v s}, Han{\v c},
  and Voz{\'a}rikov{\'a}}]{hancova_estimating_2020}
Han{\v c}ov{\'a} M, Gajdo{\v s} A, Han{\v c} J, Voz{\'a}rikov{\'a} G (2020)
  Estimating variances in time series kriging using convex optimization and
  empirical {BLUPs}. Stat Papers \doi{10.1007/s00362-020-01165-5},
  \urlprefix\url{https://doi.org/10.1007/s00362-020-01165-5}

\bibitem[{{Harrell Jr} and others(2020){Harrell Jr}
  et~al}]{harrelljr_hmisc_2020}
{Harrell Jr} FE, et~al (2020) Hmisc: {Harrell} {Miscellaneous}.
  \urlprefix\url{https://CRAN.R-project.org/package=Hmisc}

\bibitem[{Hattie(2015)}]{hattie_applicability_2015}
Hattie J (2015) The applicability of {Visible} {Learning} to higher education.
  Scholarship of Teaching and Learning in Psychology 1(1):79--91,
  \doi{10.1037/stl0000021}

\bibitem[{Hattie(2009)}]{hattie_visible_2009}
Hattie JaC (2009) Visible {Learning}: {A} {Synthesis} of {Over} 800
  {Meta}-{Analyses} {Relating} to {Achievement}. Routledge, London

\bibitem[{Haver(1999)}]{haver_calculus:_1999}
Haver WE (ed)  (1999) Calculus: {Catalyzing} a {National} {Community} for
  {Reform} : {Awards} 1987-1995. The Mathematical Association of America

\bibitem[{Heering et~al(2012)Heering, Grap{\'i}, and
  Bruneau}]{heering_innovative_2012}
Heering P, Grap{\'i} P, Bruneau O (eds)  (2012) Innovative {Methods} for
  {Science} {Education}: {History} of {Science}, {ICT} and {Inquiry} {Based}
  {Science} {Teaching}. Frank \& Timme GmbH, Berlin

\bibitem[{Heiberger and Neuwirth(2009)}]{heiberger_r_2009}
Heiberger RM, Neuwirth E (2009) R {Through} {Excel}: {A} {Spreadsheet}
  {Interface} for {Statistics}, {Data} {Analysis}, and {Graphics}. Springer,
  New York

\bibitem[{Hohenwarter et~al(2018)Hohenwarter, Borcherds, Ancsin, Bencze,
  Blossier, {\'E}li{\'a}s, Frank, G{\'a}l, Hofst{\"a}tter, Jordan, Kone{\v
  c}n{\'y}, Kov{\'a}cs, Lettner, Lizelfelner, Parisse, Solyom-Gecse, Tomaschko,
  Kuellinger, and Karacsony}]{hohenwarter_geogebra_2018}
Hohenwarter M, Borcherds M, Ancsin G, Bencze B, Blossier M, {\'E}li{\'a}s J,
  Frank K, G{\'a}l L, Hofst{\"a}tter A, Jordan F, Kone{\v c}n{\'y} Z,
  Kov{\'a}cs Z, Lettner E, Lizelfelner S, Parisse B, Solyom-Gecse C, Tomaschko
  M, Kuellinger W, Karacsony B (2018) {GeoGebra} - {Dynamic} {Mathematics} for
  {Everyone}, ver. 6.0.507.0. \urlprefix\url{http://www.geogebra.org/}

\bibitem[{Hughes-Hallett et~al(2016)Hughes-Hallett, McCallum, Gleason, Flath,
  Lock, Gordon, Lomen, Lovelock, Osgood, Pasquale, Quinney, Tecosky-Feldman,
  Thrash, Rhea, and Tucker}]{hughes-hallett_calculus:_2016}
Hughes-Hallett D, McCallum WG, Gleason AM, Flath DE, Lock PF, Gordon SP, Lomen
  DO, Lovelock D, Osgood BG, Pasquale A, Quinney D, Tecosky-Feldman J, Thrash
  J, Rhea KR, Tucker TW (2016) Calculus: {Single} and {Multivariable}, 7th edn.
  Wiley, New York

\bibitem[{Hunter(2007)}]{hunter_matplotlib_2007}
Hunter JD (2007) Matplotlib: {A} {2D} {Graphics} {Environment}. Comput Sci Eng
  9(3):90--95, \doi{10.1109/MCSE.2007.55}

\bibitem[{Van~den Hurk et~al(2016)Van~den Hurk, Houtveen, and Van~de
  Grift}]{van_den_hurk_fostering_2016}
Van~den Hurk H, Houtveen A, Van~de Grift W (2016) Fostering effective teaching
  behavior through the use of data-feedback. Teaching and Teacher Education
  60:444--451, \doi{10.1016/j.tate.2016.07.003}

\bibitem[{{IBM Corp.}(2015)}]{ibmcorp_ibm_2015}
{IBM Corp} (2015) {IBM} {SPSS} {Statistics} for {Windows}, version 23.0

\bibitem[{Ireson(1999)}]{ireson_multivariate_1999}
Ireson G (1999) A multivariate analysis of undergraduate physics students'
  conceptions of quantum phenomena. Eur J Phys 20(3):193--199,
  \doi{10.1088/0143-0807/20/3/309}

\bibitem[{Johnson and Christensen(2016)}]{johnson_educational_2016}
Johnson RB, Christensen L (2016) Educational {Research}: {Quantitative},
  {Qualitative}, and {Mixed} {Approaches}, 6th edn. SAGE Publications, London

\bibitem[{Jones et~al(2001)Jones, Oliphant, Peterson, and
  {others}}]{jones_scipy_2001}
Jones E, Oliphant T, Peterson P, {others} (2001) {SciPy}: {Open} source
  scientific tools for {Python}. \urlprefix\url{http://www.scipy.org/}

\bibitem[{kaggle(2020)}]{kaggle_kaggles_2020}
kaggle (2020) Kaggle's {State} of {Data} {Science} and {Machine} {Learning}
  2019, {Enterprise} {Executive} {Summary}.
  \urlprefix\url{https://www.kaggle.com/kaggle-survey-2019}

\bibitem[{Kassambara(2017)}]{kassambara_practical_2017}
Kassambara A (2017) Practical {Guide} to {Cluster} {Analysis} in {R}:
  {Unsupervised} {Machine} {Learning}, 1st edn. STHDA, http://www.sthda.com

\bibitem[{Kassambara and Mundt(2020)}]{kassambara_factoextra_2020}
Kassambara A, Mundt F (2020) factoextra: {Extract} and {Visualize} the
  {Results} of {Multivariate} {Data} {Analyses}.
  \urlprefix\url{https://CRAN.R-project.org/package=factoextra}

\bibitem[{Kessler(2020)}]{kessler_jupytergraffiti_2020}
Kessler W (2020) jupytergraffiti: {Create} interactive screencasts inside
  {Jupyter} {Notebook} that anybody can play back.
  \urlprefix\url{https://github.com/willkessler/jupytergraffiti}

\bibitem[{Khine and Areepattamannil(2019)}]{khine_steam_2019}
Khine MS, Areepattamannil S (2019) {STEAM} {Education}: {Theory} and
  {Practice}. Springer, Cham

\bibitem[{Kluyver et~al(2016)Kluyver, Ragan-Kelley, Perez, Granger, Bussonnier,
  Frederic, Kelley, Hamrick, Grout, Corlay, Ivanov, Avila, Abdalla, and
  Willing}]{kluyver_jupyter_2016}
Kluyver T, Ragan-Kelley B, Perez F, Granger B, Bussonnier M, Frederic J, Kelley
  K, Hamrick J, Grout J, Corlay S, Ivanov P, Avila D, Abdalla S, Willing C
  (2016) Jupyter {Notebooks}-a publishing format for reproducible computational
  workflows. In: Loizides F, Schmidt B (eds) Positioning and {Power} in
  {Academic} {Publishing}: {Players}, {Agents} and {Agendas}. {Proceedings} of
  the 20th {International} {Conference} on {Electronic} {Publishing}., Ios
  Press, Amsterdam, pp 87 -- 90

\bibitem[{Koehler and Kim(2018)}]{koehler_interactive_2018}
Koehler JF, Kim S (2018) Interactive {Classrooms} with {Jupyter} and {Python}.
  The Mathematics Teacher 111(4):304--308

\bibitem[{Kueppers(2017)}]{kueppers_latex_2017}
Kueppers B (2017) From {Latex} to {Jupyter}: {Converting} {Traditional} to
  {Modern}. In: Chova LG, Martinez AL, Torres IC (eds) Inted2017: 11th
  {International} {Technology}, {Education} and {Development} {Conference},
  Iated-Int Assoc Technology Education \& Development, Valencia, pp 1592--1596

\bibitem[{Lai and McNaughton(2016)}]{lai_impact_2016}
Lai M, McNaughton S (2016) The impact of data use professional development on
  student achievement. Teaching and Teacher Education 60:434--443,
  \doi{10.1016/j.tate.2016.07.005}

\bibitem[{LaRoche et~al(2016)LaRoche, Joncas, and Foy}]{laroche_sample_2016}
LaRoche S, Joncas M, Foy P (2016) Sample {Design} in {TIMSS} 2015. In: Martin
  MO, Mullis IV, Hooper M (eds) Methods and {Procedures} in {TIMSS} 2015,
  Boston College, TIMSS \& PIRLS International Study Center, pp 3.1--3.37

\bibitem[{Laugwitz et~al(2008)Laugwitz, Held, and
  Schrepp}]{laugwitz_construction_2008}
Laugwitz B, Held T, Schrepp M (2008) Construction and {Evaluation} of a {User}
  {Experience} {Questionnaire}. In: Holzinger A (ed) {HCI} and {Usability} for
  {Education} and {Work}, proceedings, no. 5298 in Lecture {Notes} in
  {Computer} {Science}, Springer Berlin Heidelberg, pp 63--76

\bibitem[{Lohr(2009)}]{lohr_sampling_2009}
Lohr SL (2009) Sampling: {Design} and {Analysis}. Cengage Learning, Boston

\bibitem[{L{\"u}decke(2020)}]{ludecke_sjstats_2020}
L{\"u}decke D (2020) sjstats: {Collection} of {Convenient} {Functions} for
  {Common} {Statistical} {Computations}.
  \urlprefix\url{https://CRAN.R-project.org/package=sjstats}

\bibitem[{Maechler et~al(2019)Maechler, original), original), original),
  Hornik~[trl, maintenance(1999 2000)), Studer, Roudier, Gonzalez, Kozlowski,
  pam()), and Murphy~(volume.ellipsoid(\{d {\textgreater}=
  3\}))}]{maechler_cluster_2019}
Maechler M, original) PRF, original) ASS, original) MHS, Hornik~[trl K,
  maintenance(1999 2000)) cptR, Studer M, Roudier P, Gonzalez J, Kozlowski K,
  pam()) ESfof, Murphy~(volumeellipsoid(\{d {\textgreater}= 3\})) K (2019)
  cluster: "{Finding} {Groups} in {Data}": {Cluster} {Analysis} {Extended}
  {Rousseeuw} et al. \urlprefix\url{https://CRAN.R-project.org/package=cluster}

\bibitem[{Mandinach(2016)}]{mandinach_teachers_2016}
Mandinach EB (2016) Teachers learning how to use data: {A} synthesis of the
  issues and what is known. Teaching and Teacher Education p~6

\bibitem[{Mandinach and Gummer(2016)}]{mandinach_what_2016}
Mandinach EB, Gummer ES (2016) What does it mean for teachers to be data
  literate: {Laying} out the skills, knowledge, and dispositions. Teaching and
  Teacher Education 60:366--376, \doi{10.1016/j.tate.2016.07.011}

\bibitem[{Marr(2018)}]{marr_how_2018}
Marr B (2018) How much data do we create every day? {The} mind-blowing stats
  everyone should read.
  \urlprefix\url{https://www.forbes.com/sites/bernardmarr/2018/05/21/how-much-data-do-we-create-every-day-the-mind-blowing-stats-everyone-should-read}

\bibitem[{Maxwell et~al(2017)Maxwell, Delaney, and
  Kelley}]{maxwell_designing_2017}
Maxwell SE, Delaney HD, Kelley K (2017) Designing {Experiments} and {Analyzing}
  {Data}: {A} {Model} {Comparison} {Perspective}, 3rd edn. Routledge

\bibitem[{Mayer(2008)}]{mayer_applying_2008}
Mayer RE (2008) Applying the science of learning: {Evidence}-based principles
  for the design of multimedia instruction. Am Psychol 63(8):760--769

\bibitem[{Mayer(2014{\natexlab{a}})}]{mayer_cambridge_2014}
Mayer RE (ed)  (2014{\natexlab{a}}) The {Cambridge} {Handbook} of {Multimedia}
  {Learning}, 2nd edn. Cambridge University Press, Cambridge

\bibitem[{Mayer(2014{\natexlab{b}})}]{mayer_cognitive_2014}
Mayer RE (2014{\natexlab{b}}) Cognitive {Theory} of {Multimedia} {Learning},
  chap. 3. In: Mayer RE (ed) The {Cambridge} {Handbook} of {Multimedia}
  {Learning}, 2nd edn, Cambridge University Press, Cambridge, pp 67--108

\bibitem[{Mayer(2019)}]{mayer_thirty_2019}
Mayer RE (2019) Thirty years of research on online learning. Applied Cognitive
  Psychology 33(2):152--159, \doi{10.1002/acp.3482}

\bibitem[{Mazur and Watkins(2009)}]{simkins_just--time_2009}
Mazur E, Watkins J (2009) Just-in-{Time} {Teaching} and {Peer} {Instruction},
  chap. 3. In: Simkins S, Maier M (eds) Just in {Time} {Teaching}: {Across} the
  {Disciplines}, and {Across} the {Academy}, Stylus Publishing, Sterling, pp
  39--62

\bibitem[{McKagan et~al(2016)McKagan, Madsen, Barbato, Mason, Sayre,
  Cunningham, Hilborn, Riggsbee, Martinuk, and Bell}]{mckagan_physport:_2016}
McKagan S, Madsen A, Barbato L, Mason B, Sayre E, Cunningham B, Hilborn B,
  Riggsbee M, Martinuk S, Bell A (2016) Physport: {Supporting} physics teaching
  with research-based resources. \urlprefix\url{https://www.physport.org/}

\bibitem[{McKagan et~al(2019)McKagan, Strubbe, Barbato, Madsen, Sayre, and
  Mason}]{mckagan_physport_2019}
McKagan SB, Strubbe LE, Barbato LJ, Madsen AM, Sayre EC, Mason BA (2019)
  {PhysPort} use and growth: {Supporting} physics teaching with research-based
  resources since 2011. arXiv:190503745 [physics] ArXiv: 1905.03745

\bibitem[{McKiernan(2017)}]{mckiernan_imagining_2017}
McKiernan EC (2017) Imagining the {\textquotedblleft}open{\textquotedblright}
  university: {Sharing} scholarship to improve research and education. PLoS
  Biol 15(10):e1002,614, \doi{10.1371/journal.pbio.1002614}

\bibitem[{McKinney(2010)}]{mckinney_data_2010}
McKinney W (2010) Data {Structures} for {Statistical} {Computing} in {Python}.
  In: Walt Svd, Millman J (eds) Proceedings of the 9th {Python} in {Science}
  {Conference}, pp 51 -- 56

\bibitem[{Ndukwe and Daniel(2020)}]{ndukwe_teaching_2020}
Ndukwe IG, Daniel BK (2020) Teaching analytics, value and tools for teacher
  data literacy: a systematic and tripartite approach. Int J Educ Technol High
  Educ 17(1):22, \doi{10.1186/s41239-020-00201-6}

\bibitem[{Nouri(2016)}]{nouri_flipped_2016}
Nouri J (2016) The flipped classroom: for active, effective and increased
  learning {\textendash} especially for low achievers. Int J Educ Technol High
  Educ 13(1):33, \doi{10.1186/s41239-016-0032-z}

\bibitem[{Odden and Caballero(2019)}]{odden_computational_2019}
Odden TOB, Caballero MD (2019) Computational {Essays}: {An} {Avenue} for
  {Scientific} {Creativity} in {Physics}. In: Cao Y, Wolf S, Bennet M (eds)
  2019 {Physics} {Education} {Research} {Conference} {Proceedings} [{Provo},
  {UT}, {July} 27-25, 2019, \doi{doi:10.1119/perc.2019.pr.Odden}, available
  also as arXiv: 1909.12697

\bibitem[{Odden et~al(2019)Odden, Lockwood, and Caballero}]{odden_physics_2019}
Odden TOB, Lockwood E, Caballero MD (2019) Physics computational literacy: {An}
  exploratory case study using computational essays. Phys Rev Phys Educ Res
  15(2):020,152, \doi{10.1103/PhysRevPhysEducRes.15.020152}

\bibitem[{Oliphant(2007)}]{oliphant_python_2007}
Oliphant TE (2007) Python for {Scientific} {Computing}. Comput Sci Eng
  9(3):10--20

\bibitem[{Panko(1998)}]{panko_what_1998}
Panko RR (1998) What {We} {Know} {About} {Spreadsheet} {Errors}. Journal of End
  User Computing's 10(Special issue on Scaling Up End User Development):15--21

\bibitem[{Pa{\v n}kov{\'a} and Han{\v
  c}(2019{\natexlab{a}})}]{pankova_flipped_2019}
Pa{\v n}kov{\'a} E, Han{\v c} J (2019{\natexlab{a}}) Flipped learning and
  interactive methods with smartphones in modern physics at secondary schools.
  AIP Conference Proceedings 2152(1):030,025, \doi{10.1063/1.5124769}

\bibitem[{Pa{\v n}kov{\'a} and Han{\v
  c}(2019{\natexlab{b}})}]{pankova_teaching_2019}
Pa{\v n}kov{\'a} E, Han{\v c} J (2019{\natexlab{b}}) Teaching
  {Feynman}{\textquoteright}s quantum physics at secondary schools using
  current digital technologies. AIP Conference Proceedings 2152(1):030,026,
  \doi{10.1063/1.5124770}

\bibitem[{Pa{\v n}kov{\'a} et~al(2016)Pa{\v n}kov{\'a}, {\v S}trauch, and
  Han{\v c}}]{pankova_practical_2016}
Pa{\v n}kov{\'a} E, {\v S}trauch P, Han{\v c} J (2016) Practical strategies in
  formative and summative assessment of the flipped math and physics education.
  In: Santiago~Campi{\'o}n R (ed) Actas del {II} {Congreso} de {Flipped}
  {Classroom}: {Comunicaciones} y posters presentados, Edita MT Servicios
  Educativos, Zaragoza, pp 308--327

\bibitem[{Poortman et~al(2016)Poortman, Schildkamp, and
  Lai}]{poortman_professional_2016}
Poortman CL, Schildkamp K, Lai MK (2016) Professional development in data use:
  {An} international perspective on conditions, models, and effects. Teaching
  and Teacher Education 60:363--365, \doi{10.1016/j.tate.2016.07.029}

\bibitem[{{Project Jupyter} et~al(2018){Project Jupyter}, Bussonnier, Forde,
  Freeman, Granger, Head, Holdgraf, Kelley, Nalvarte, Osheroff, Pacer, Panda,
  Perez, Ragan-Kelley, and Willing}]{projectjupyter_binder_2018}
{Project Jupyter}, Bussonnier M, Forde J, Freeman J, Granger B, Head T,
  Holdgraf C, Kelley K, Nalvarte G, Osheroff A, Pacer M, Panda Y, Perez F,
  Ragan-Kelley B, Willing C (2018) Binder 2.0 - {Reproducible}, interactive,
  sharable environments for science at scale. Proceedings of the 17th Python in
  Science Conference pp 113--120

\bibitem[{{R Development Core Team}(2020)}]{r_development_core_team_r:_2020}
{R Development Core Team} (2020) R: {A} language and environment for
  statistical computing. \urlprefix\url{http://www.R-project.org/}

\bibitem[{Redish(2014)}]{redish_oersted_2014}
Redish EF (2014) Oersted {Lecture} 2013: {How} should we think about how our
  students think? American Journal of Physics 82(6):537--551,
  \doi{10.1119/1.4874260}

\bibitem[{Romer(2018)}]{romer_jupyter_2018}
Romer P (2018) Jupyter, {Mathematica}, and the {Future} of the {Research}
  {Paper}.
  \urlprefix\url{https://paulromer.net/jupyter-mathematica-and-the-future-of-the-research-paper/}

\bibitem[{Salkind(2010)}]{salkind_encyclopedia_2010}
Salkind NJ (ed)  (2010) Encyclopedia of {Research} {Design}, 1st edn. SAGE
  Publications, Inc

\bibitem[{Schrepp et~al(2017)Schrepp, Hinderks, and
  Thomaschewski}]{schrepp_design_2017}
Schrepp M, Hinderks A, Thomaschewski J (2017) Design and {Evaluation} of a
  {Short} {Version} of the {User} {Experience} {Questionnaire} ({UEQ}-{S}).
  International Journal of Interactive Multimedia and Artificial Intelligence
  4(6):103--108

\bibitem[{Sergis et~al(2019)Sergis, Sampson, Rodr{\'i}guez-Triana, Gillet,
  Pelliccione, and de~Jong}]{sergis_using_2019}
Sergis S, Sampson DG, Rodr{\'i}guez-Triana MJ, Gillet D, Pelliccione L, de~Jong
  T (2019) Using educational data from teaching and learning to inform
  teachers{\textquoteright} reflective educational design in inquiry-based
  {STEM} education. Computers in Human Behavior 92:724--738,
  \doi{10.1016/j.chb.2017.12.014}

\bibitem[{Somers(2018)}]{somers_scientific_2018}
Somers J (2018) The {Scientific} {Paper} {Is} {Obsolete}. The Atlantic
  \urlprefix\url{https://www.theatlantic.com/science/archive/2018/04/the-scientific-paper-is-obsolete/556676/}

\bibitem[{Spector(2020)}]{spector_bringing_2020}
Spector C (2020) Bringing math class into the data age.
  \urlprefix\url{https://ed.stanford.edu/news/bringing-math-class-data-age}

\bibitem[{Springuel et~al(2019)Springuel, Wittmann, and
  Thompson}]{springuel_reconsidering_2019}
Springuel RP, Wittmann MC, Thompson JR (2019) Reconsidering the encoding of
  data in physics education research. Phys Rev Phys Educ Res 15(2):020,103,
  \doi{10.1103/PhysRevPhysEducRes.15.020103}

\bibitem[{Stein and {others}(2020)}]{stein_sage_2020}
Stein WA, {others} (2020) Sage {Mathematics} {Software}, ver. 9.1.
  \urlprefix\url{http://www.sagemath.org}

\bibitem[{{\v S}trauch and Han{\v c}(2017)}]{strauch_quantitative_2017}
{\v S}trauch P, Han{\v c} J (2017) Quantitative diagnostics of misconceptions
  in science education [in {Slovak}]. Eduk{\'a}cia 2(2):11 pp.,
  \urlprefix\url{https://www.upjs.sk/public/media/15903/Strauch_Hanc.pdf}

\bibitem[{Talbert and Bergmann(2017)}]{talbert_flipped_2017}
Talbert R, Bergmann J (2017) Flipped {Learning}: {A} {Guide} for {Higher}
  {Education} {Faculty}. Stylus Publishing, Sterling, Virginia

\bibitem[{VanderPlas(2016)}]{vanderplas_python_2016}
VanderPlas J (2016) Python {Data} {Science} {Handbook}: {Essential} {Tools} for
  {Working} with {Data}. O'Reilly Media, Inc., Boston

\bibitem[{Walt et~al(2011)Walt, Colbert, and Varoquaux}]{walt_numpy_2011}
Walt Svd, Colbert SC, Varoquaux G (2011) The {NumPy} {Array}: {A} {Structure}
  for {Efficient} {Numerical} {Computation}. Comput Sci Eng 13(2):22--30,
  \doi{10.1109/MCSE.2011.37}

\bibitem[{Warner(2018)}]{warner_thank_2018}
Warner J (2018) Thank you for 100 million repositories.
  \urlprefix\url{https://github.blog/2018-11-08-100m-repos/}

\bibitem[{Weiss(2017)}]{weiss_scientific_2017}
Weiss CJ (2017) Scientific {Computing} for {Chemists}: {An} {Undergraduate}
  {Course} in {Simulations}, {Data} {Processing}, and {Visualization}. J Chem
  Educ 94(5):592--597, \doi{10.1021/acs.jchemed.7b00078}

\bibitem[{Wickham et~al(2019)Wickham, Bryan, attribution), code), code), code),
  code), code), and code)}]{wickham_readxl_2019}
Wickham H, Bryan J, attribution) RChoaRcaaCcwec, code) MKAoiR, code) KVAoil,
  code) CLAoil, code) BCAoil, code) DHAoil, code) EMAoil (2019) readxl: {Read}
  {Excel} {Files}. \urlprefix\url{https://CRAN.R-project.org/package=readxl}

\bibitem[{Wickham et~al(2020{\natexlab{a}})Wickham, Fran{\c c}ois, Henry,
  M{\"u}ller, and RStudio}]{wickham_dplyr_2020}
Wickham H, Fran{\c c}ois R, Henry L, M{\"u}ller K, RStudio (2020{\natexlab{a}})
  dplyr: {A} {Grammar} of {Data} {Manipulation}.
  \urlprefix\url{https://CRAN.R-project.org/package=dplyr}

\bibitem[{Wickham et~al(2020{\natexlab{b}})Wickham, Seidel, and
  RStudio}]{wickham_scales_2020}
Wickham H, Seidel D, RStudio (2020{\natexlab{b}}) scales: {Scale} {Functions}
  for {Visualization}.
  \urlprefix\url{https://CRAN.R-project.org/package=scales}

\bibitem[{Wilcox(2017)}]{wilcox_understanding_2017}
Wilcox RR (2017) Understanding and {Applying} {Basic} {Statistical} {Methods}
  {Using} {R}, 1st edn. Wiley, New York

\bibitem[{Wright et~al(2020)Wright, Schwartz, Oaks, Newman, and
  Flanagan}]{wright_why_2020}
Wright AM, Schwartz RS, Oaks JR, Newman CE, Flanagan SP (2020) The why, when,
  and how of computing in biology classrooms. F1000Res 8,
  \doi{10.12688/f1000research.20873.2},
  \urlprefix\url{https://www.ncbi.nlm.nih.gov/pmc/articles/PMC6971840/}

\bibitem[{Zaiontz(2019)}]{zaiontz_real_2019}
Zaiontz C (2019) Real {Statistics} {Resource} {Pack} {\textbar} {Real}
  {Statistics} {Using} {Excel}. \urlprefix\url{http://www.real-statistics.com}

\bibitem[{van~der Zee and Reich(2018)}]{van_der_zee_open_2018}
van~der Zee T, Reich J (2018) Open {Education} {Science}. AERA Open
  4(3):233285841878,746, \doi{10.1177/2332858418787466}

\bibitem[{Zimmermann et~al(2018)Zimmermann, Casamayou, Cohen, Connan, Dumont,
  Fousse, Maltey, Meulien, Mezzarobba, Pernet, Thi{\'e}ry, Bray, Cremona,
  Forets, Ghitza, and Thomas}]{zimmermann_computational_2018}
Zimmermann P, Casamayou A, Cohen N, Connan G, Dumont T, Fousse L, Maltey F,
  Meulien M, Mezzarobba M, Pernet C, Thi{\'e}ry NM, Bray E, Cremona J, Forets
  M, Ghitza A, Thomas H (2018) Computational {Mathematics} with {SageMath}.
  SIAM, Philadelphia

\end{thebibliography}

\newpage
\section*{Appendix}

\subsection*{Study I: Two open questions of the qualitative interview}
\begin{itemize}
	\item \textit{Explain personal reasons of your choosing statistical methods  and digital tools.}
	\item \textit{Describe key obstacles why data science tool R with R Commander did not fit your research needs and requirements.}
\end{itemize}

\subsection*{Study II: Digital Experience Questionnaire}
\begin{itemize}
	\item \textit{What mobile device WIFI connectable to the Internet you can bring to our course, if we were not in a PC room?} 
	\item \textit{What device you can use in home preparation, i.e. to study and view the learning e-content (course website, videos, presentations, interactive documents, simulations)?}
	\item \textit{From which secondary school do you come?}
	\item \textit{The most frequent half-year or end-year final grade from secondary school math: }
	\item \textit{What digital tools did you regularly use in math lessons at your secondary school?}
\end{itemize}

\subsection*{Study II: Three additional questions in the UEQ.}
The complete 26 items of UEQ questionnaire in English can be found at  \href{https://www.ueq-online.org/}{ueq-online.org}.

\begin{itemize}
	\item \textit{The use of digital technology has helped me a lot to understand ideas we have learned.}
	\item \textit{I have greatly improved in the digital technology we used in learning.}
	\item \textit{The technology would also be very helpful in other subjects.}
\end{itemize}
\vspace{6pt}

\subsection*{Study III: Teacher's Attitude Questionnaire.}
The complete Slovak and English version of our short semi-closed, self-developed questionnaire is available at our GitHub storage. 

\begin{itemize}
	\item \textit{Future app use in own}: Would you like to use our R shiny app in the future? 
	\item \textit{Next app extension to more tests and subjects:} How do you perceive further extension of our web app?
\end{itemize}
\vspace{6pt}

\section*{Additional Files}

Our open data analysis in the form of Jupyter notebooks together with all used tools and data files are stored and freely available at one of our GitHub repository devoted to this paper \citep{hanc_jupyterper_2020-1} in the frame of our GitHub research project \textit{Jupyter in Physics education and Research} \citep{hanc_jupyterper_2020}.

\end{backmatter}
\end{document}